\documentclass[ALICE,manyauthors]{cernphprep}
\usepackage[english]{babel}
\usepackage{graphicx}
\usepackage{verbatim}
\usepackage{amsfonts}
\usepackage{makeidx}
\usepackage{color}
\usepackage{amsmath}
\usepackage{commath}
\usepackage{lineno}
\usepackage{epsfig}
\usepackage{epstopdf}
\usepackage{hyperref}
\usepackage{cite}
\usepackage{multirow}
\usepackage{subfigure}
\usepackage{placeins}
\newcommand{\sqrts}{\sqrt{s}}
\newcommand{\sqrtsNN}{\sqrt{s_{\scriptscriptstyle \rm NN}}}

\newcommand{\gev}{\mathrm{GeV}}
\newcommand{\gevc}{\mathrm{GeV}/c}
\newcommand{\tev}{\mathrm{TeV}}

\newcommand{\mm}{\mathrm{mm}}
\newcommand{\cm}{\mathrm{cm}}

\newcommand{\pt}{p_{\rm T}}
\newcommand{\ptD}{p_{\rm T}^{\rm D}}
\newcommand{\ptAssoc}{p_{\rm T}^{\rm assoc}}

\newcommand{\DtoKpi}{{\rm D}^0 \to {\rm K}^-\pi^+}
\newcommand{\DtoKpipi}{{\rm D}^+\to {\rm K}^-\pi^+\pi^+}
\newcommand{\DstartoDpi}{{\rm D}^{*+} \to {\rm D}^0 \pi^+}

\newcommand{\Dzero}{{\rm D^0}}

\newcommand{\Dstar}{{\rm D^{*+}}}

\newcommand{\Dplus}{{\rm D^+}}

\newcommand{\ccbar}{{\rm c}\bar{\rm c}}
\newcommand{\bbbar}{{\rm b}\bar{\rm b}}
\newcommand{\nbinv}{{\rm nb^{-1}}}

\newcommand{\pPb}{\mbox{p--Pb}}
\newcommand{\dphi}{\Delta \mathrm{\varphi}}
\newcommand{\deta}{\Delta \mathrm{\eta}}
\newcommand{\vtwo}{v_{2}} 
\newcommand{\pythia}{\rm PYTHIA}

\newcommand{\powheg}{\rm POWHEG}

\begin{document}%

\begin{titlepage}
\PHyear{2016}
\PHnumber{129}      
\PHdate{20 May}  
%

\title{Measurement of azimuthal correlations of D mesons with charged particles in pp collisions at $\mathbf{\sqrt{s}=7}$~$\mathbf{\tev}$ and p--Pb collisions at $\mathbf{\sqrtsNN=5.02~\tev}$ }
\ShortTitle{D-meson with charged-particle azimuthal correlations in pp and p--Pb collisions}   

\Collaboration{ALICE Collaboration\thanks{See Appendix~\ref{app:collab} for the list of collaboration members}}
\ShortAuthor{ALICE Collaboration} 

\begin{abstract}
The azimuthal correlations of D mesons with charged particles were measured with the ALICE apparatus in pp collisions
at $\sqrts=7~\tev$ and p--Pb collisions at $\sqrtsNN=5.02~\tev$ at the Large Hadron Collider. $\Dzero$, $\Dplus$, and $\Dstar$
mesons and their charge conjugates with transverse momentum $3<\pt<16~\gevc$ and rapidity in the nucleon-nucleon centre-of-mass system $|y_{\rm cms}|<0.5$ (pp collisions)
and $-0.96<y_{\rm cms}<0.04$ (p--Pb collisions) were correlated to charged particles with $\pt>0.3~\gevc$. The yield of charged
particles in the correlation peak induced by the jet containing the D meson and the peak width are compatible within uncertainties in 
the two collision systems. The data are described
within uncertainties by Monte-Carlo simulations based on PYTHIA, POWHEG, and EPOS~3 event generators.

\end{abstract}
\end{titlepage}
\setcounter{page}{2}
%
%
\section{Introduction}
\label{sec:intro}
The study of the angular correlation of D mesons with charged particles, i.e. the distribution of 
the differences in azimuthal angles, $\dphi=\varphi_{\rm ch}-\varphi_{\rm D}$, and pseudorapidities, $\Delta\eta=\eta_{\rm ch}-\eta_{\rm D}$, 
allows for the characterisation of charm production and fragmentation processes in proton--proton (pp) collisions
and of their possible modifications due to nuclear effects in proton--Pb and Pb--Pb collisions~\cite{Beraudo2014DhCorrelPbPb,beraudoHFv2pPb}. For
leading-order (LO) Quantum-ChromoDynamic (QCD) processes, charm quark-antiquark pairs are produced back-to-back in azimuth: the angular correlation
of D mesons with charged particles features a ``near-side'' peak at $(\dphi,\deta)=(0,0)$ and an ``away-side'' peak  at $\dphi=\pi$. The former
originates from the jet containing the ``trigger'' D meson, the latter is induced by the recoil jet, which 
can also include the decay products of the other charmed hadron produced in the collision. The away-side peak extends over a wide range
in $\Delta\eta$. The two peaks lie on top of an approximately flat distribution
arising from the correlation of D mesons with charged particles from the underlying event. %
Next-to-leading order (NLO) production processes can give rise to significantly different correlation 
patterns~\cite{mangano,pythiaHQ}. For example, the radiation of a hard
gluon from a charm quark smears the
back-to-back topology of LO production and broadens both the near- and the away-side peak. In addition, quark-antiquark charm pairs originating from the 
splitting of a gluon can be
rather collimated and, especially at high transverse momentum ($\pt$), can generate sprays of hadrons
contributing to a unique and broader ``near-side'' peak. In such cases, the away-side peak stems from the particles coming from the fragmentation of the recoil parton 
(typically a gluon or a light quark), which is not aligned with the trigger D meson. Finally, in hard-scattering topologies classified 
as ``flavour excitation'' (see e.g.~\cite{pythiaHQ}) a charm quark (antiquark) from an initial splitting $g\rightarrow c\bar{c}$ undergoes a 
hard interaction. The hadrons originating 
from the antiquark (quark) can be significantly separated in rapidity with respect to the trigger D meson and  
contribute with a rather flat term to the $\dphi$ distribution. %

Since the first measurement performed by 
STAR in Au--Au collisions at $\sqrtsNN=200~\gev$~\cite{IaaSTAR2002}, two-particle azimuthal correlations have been exploited 
at both RHIC and the LHC~\cite{IaaSTAR,IaaPHENIX,IaaALICE} to investigate the possible modifications of jet and dijet 
properties that can be caused by the interaction of high-energy partons
with the constituents of the Quark Gluon Plasma (QGP) formed in ultra-relativistic heavy-ion collisions. The most evident effect is the 
suppression of the away-side
correlation peak, commonly attributed to in-medium partonic energy loss. The results allow one to constrain the dependence of the energy loss on the distance covered
by partons in the QGP as well as the initial gluon density~\cite{Zhang:2007ja,Renk:2007id}. The correlation pattern of hadron-hadron pairs primarily arises from the back-to-back production of gluons or light-quarks
produced in hard-scattering processes, and their subsequent fragmentation. PHENIX measured the azimuthal 
correlation of electrons from heavy-flavour hadron decays
with charged particles in Au--Au collisions at $\sqrtsNN=200~\gev$~\cite{PHENIXhfehadAuAu}. The near- and away-side peaks are 
suppressed by factors compatible, within
uncertainties, to those observed for hadron-hadron correlations, if a similar $\pt$ is considered for the trigger hadron and the electron parent hadron. %
The proper 
interpretation of nucleus-nucleus results and the connection of the modifications of the correlation peak properties to the parton dynamics in 
the QGP requires the comparison of data with model predictions. It is crucial that the models reproduce the correlation
pattern measured in pp collisions, where nuclear effects are absent, as well as the production spectra in both pp and nucleus--nucleus collisions.
Therefore, the measurement of azimuthal correlations of D mesons with charged particles
in pp and p--Pb collisions serves not only as a reference for future measurements in Pb-Pb collisions but it also allows for validation of
Monte-Carlo generator expectations, which is fundamental for the understanding of the results in all collision systems.

Perturbative QCD calculations relying on the collinear-factorisation approach, like
FONLL~\cite{fonll} and GM-VFNS~\cite{gmvfns}, or based on the $k_{\rm T}$-factorisation approach~\cite{ktfact}
describe reasonably well the $\pt$-differential production
cross sections of D mesons from charm-quark fragmentation measured at central rapidity ($|y|<0.5$)
in pp collisions at $\sqrt{s}=7$ and $2.76~\tev$  using
the ALICE detector~\cite{DmesonPP,DmesonPP276}. These calculations represent the state of the art for the 
computation of ($\pt,y$)-differential cross sections of charm quarks and charmed hadrons. However, the kinematic relationship between D mesons and 
particles from charm fragmentation and the underlying event is accessible only with event generators 
coupled with parton-shower Monte-Carlo programs like those provided by PYTHIA~\cite{pythia} and HERWIG~\cite{herwig}. The order of
hard-scattering matrix elements used, the specific implementation of parton shower and hadronisation, as well as the modeling of the
underlying event have an influence on the angular correlations of D mesons with charged particles produced in the event. For heavy quarks with mass $M$ 
and energy $E_{\rm Q}$, the suppression of gluon radiation off the quark inside the forward cone 
with opening angle $\Theta=M/E_{\rm Q}$ (the so-called ``dead-cone'' effect) reduces the phase space for primary gluon radiation~\cite{Dokshitzer1991}. This implies a harder fragmentation of 
the quarks into the heavy hadrons and leads to essential
differences in the profiles of gluon-, light-quark- and heavy-quark-initiated jets resulting in shape differences
of $\pt$-spectra and multiplicity distributions of primary hadrons in the jets~\cite{Dokshitzer2006,Ramos2010}.

Correlations between D mesons were measured at the LHC in pp collisions
at $\sqrts=7~\tev$ with the LHCb experiment~\cite{LHCbcharm}, providing information on charm production mechanisms and on the
properties of events containing heavy flavours. ATLAS measured the production of $\Dstar$ mesons in jets in pp collisions
at $\sqrts=7~\tev$ for jets with $25<\pt<70~\gev/c$ and $\Dstar$ carrying a jet momentum fraction ($z$) in the range $0.3<z<1$. The results
indicate that the production of charm-quark jets or charm-quark fragmentation into $\Dstar$ mesons is not properly modeled
in state-of-the-art Monte-Carlo generators~\cite{ATLASDjets}. Azimuthal correlations of electrons from heavy-flavour hadron decays with charged particles
were also exploited to study the relative beauty contribution to the population of electrons from heavy-flavour hadron decays in pp
collisions at RHIC and at the LHC~\cite{STAReh,ALICEeh}. %

The angular distribution of particles produced in an event is sensitive to collective effects that correlate particle production over
wide phase-space regions. This is particularly relevant in Pb--Pb collisions with non-zero collision impact parameter, where the azimuthal asymmetry
of the overlapping region of the colliding nuclei gives rise to anisotropic pressure gradients inducing an anisotropy in the azimuthal distribution of particle momenta~\cite{PbPbv2ALICE,PbPbvnALICE}. %
The main component of the Fourier decomposition used to describe the resulting $\dphi$ distribution of two particle correlations is
the $2^{\rm nd}$ order term, proportional to $\cos(2\dphi)$, called elliptic flow or $\vtwo$. Given that correlations induced by the collective
motion of the system extend over large pseudorapidity ranges, the elliptic-flow term manifests itself with the
presence of two long-range ridge-like structures in the near and away sides of two-particle angular correlations. Unexpectedly, similar 
long-range correlation structures were observed in high-multiplicity pp and p--Pb collisions at the 
LHC~\cite{ppCMS,ppATLAS,pPbdoubleridgeALICE,pPbmuonTracklCorrelALICE,pPbdoubleridgeATLAS,pPbdoubleridgeCMS}. Also in central d--Au collisions at RHIC~\cite{dAuPHENIX,dAuSTARlongrange} similar results were obtained, although contributions from jet-like
correlations due to biases on the event selection could not be excluded~\cite{dAuSTAR}. The origin of such $\vtwo$-like structures is still
debated. Positive $\vtwo$ values in high-multiplicity pp collisions and p--Pb (d--Au) collisions at LHC (RHIC) are
expected in models that include final-state effects~\cite{werner, Alderweireldt,Bozek:2013uha,Bozek:2012gr,He:2015hfa}, as well as initial-state 
effects related to the Color Glass Condensate~\cite{dusling} or to gluon bremsstrahlung by a quark-antiquark string~\cite{arbuzov}. A modification of the azimuthal correlations of D mesons with charged particles in p--Pb
with respect to pp collisions could be a signal of the presence of long-range $\vtwo$-like correlations for particles originating from
hard-scattering processes. This would yield complementary information to that obtained
from correlations of light-flavour particles, which at low $\pt$ are primarily produced in soft processes. The
D-meson $\pt$-differential production cross section in p--Pb collisions at $\sqrtsNN=5.02~\tev$ was measured
with ALICE in the interval of rapidity in the nucleon-nucleon centre-of-mass system $-0.96<y_{\rm cms}<0.04$~\cite{ALICEDmesonpPb}. The data are
compatible, within uncertainties, with a Glauber-model-based geometrical scaling of a pp collision reference obtained from
the cross sections measured at $\sqrts=7~\tev$ and $\sqrts=2.76~\tev$. This suggests that nuclear effects are
rather small for D mesons in the range $1<\pt<24~\gevc$. However, they could still affect
angular correlations as observed at RHIC for
azimuthally-correlated pairs of electrons and muons from decays of heavy-flavour hadrons in d--Au
collisions at $\sqrtsNN=200~\gev$~\cite{PHENIXemudAu}. A
modification of the azimuthal correlation of heavy-flavour particles in p--Pb collisions could occur 
at the LHC due to gluon saturation in the heavy nucleus~\cite{fujiiwatanabe}. Moreover, transport
models based on the Langevin equation~\cite{beraudoHFv2pPb,dukev2pPb} describe, within uncertainties, the
nuclear modification factor of D mesons measured in p--Pb collisions at the LHC and
that of electrons from heavy-flavour hadron decays measured in d--Au collisions at RHIC~\cite{PHENIXdAuHFe}. These models
assume the formation of a small-size QGP in p--Pb and d--Au collisions
and include the possibility of heavy-flavour hadron formation via coalescence of heavy quarks with thermalised light quarks
from the medium. These transport calculations predict a
positive D-meson $\vtwo$ in central p--Pb collisions. As an example, in the case of the POWLANG
model~\cite{beraudoHFv2pPb} the maximum expectation for the 20\% most central p--Pb collisions is $\vtwo\sim5\%$ at $\pt=4~\gev/c$. A finite $\vtwo$ of muons from heavy-flavour hadron decays in high-multiplicity p--Pb collisions
was also suggested in~\cite{pPbmuonTracklCorrelALICE} as one of the possibilities for reconciling the measured values of
$\vtwo$ of inclusive muons with the expectations based on the multi-phase transport model AMPT~\cite{AMPTpPb}.

In this paper we report the first measurements of azimuthal correlations of D mesons with charged primary 
particles in pp and p--Pb collisions at $\sqrts=7~\tev$ and $\sqrtsNN=5.02~\tev$, respectively. Unless 
differently specified we always refer to ``prompt'' D mesons from charm-quark fragmentation. In what follows, primary
particles are defined as particles originated at the collision point, including those deriving from strong
and electromagnetic decays of unstable particles, and those from decays of hadrons with charm or beauty. The paper is organised
as follows. In Section~\ref{sec:detector} the data samples used and the details of the ALICE experimental apparatus
relevant for this analysis are described. The analysis strategy, the D-meson signal extraction, the associated-track selection criteria,
and the corrections applied to measure the correlations between D mesons and charged primary particles are
reported in Section~\ref{sec:Analysis}. In the same section, the fit procedure adopted to quantify the correlation peak properties
is described. Section~\ref{sec:systematics} reports the systematic uncertainties affecting the measurement. The results
are discussed in Section~\ref{sec:Results}. The paper is then summarised in Section~\ref{sec:conclusions}. 
\section{Experimental apparatus and data samples}
\label{sec:detector}
\subsection{The ALICE detector and event selection}
The ALICE apparatus~\cite{aliceJINST,alicePerf2014} consists of a central barrel embedded in a 0.5~T solenoidal magnetic field, a forward
muon spectrometer, and a set of
detectors located in the forward- and backward-rapidity regions dedicated to trigger and event characterisation. %
The analysis reported in this paper is performed using the central barrel detectors. Charged particle tracks are reconstructed using the Inner Tracking
System (ITS), consisting of
six layers of silicon detectors, and the Time Projection Chamber (TPC). Particle identification (PID) is based on the
specific energy loss d$E$/d$x$ in the TPC gas and on the time of flight from the interaction vertex to the Time-Of-Flight~(TOF)
detector. The ITS, TPC and TOF detectors
provide full azimuthal coverage in the pseudorapidity interval $|\eta|<0.9$.

The pp data sample consists of about $3\cdot10^8$  minimum-bias events, corresponding to an
integrated luminosity of $L_{\rm int} = 5~\nbinv$.  These collisions are triggered by the presence of at least 
one hit in one of the V0 scintillator arrays, covering the ranges $-3.7<\eta<-1.7$ and $2.8<\eta<5.1$, or in the
Silicon Pixel Detector (SPD), constituting the two innermost layers of the ITS, with an acceptance of $|\eta|<2$ (inner layer) and $|\eta|<1.4$ (outer layer).
The p--Pb data sample consists of about $10^8$ minimum-bias events, corresponding to an
integrated luminosity of about $L_{\rm int}=50~{\rm \mu b^{-1}}$. In this case the minimum-bias trigger
requires signals in both the V0 detectors. 

Only events with a reconstructed primary vertex within $\pm 10$~cm from the centre of the detector along the beam line are considered for both pp and p--Pb collisions.  This choice maximises the detector coverage of the selected events, considering the longitudinal size of the interaction region, and the detector
pseudorapidity acceptances (for more details see~\cite{alicePerf2014}). For p--Pb collisions, the center-of-mass reference
frame of the nucleon-nucleon collision is shifted in rapidity by $\Delta y_{\rm{NN}} = 0.465$ in the proton direction with respect to the laboratory frame, due 
to the different per-nucleon energies of the proton and the lead beams.

Beam-gas events are removed by offline selections based on the timing information provided
by the V0 and the Zero Degree Calorimeters (two sets of neutron and proton calorimeters located around 110~m from the
interaction point along the beam direction), and the correlation between the number of hits and track segments in the SPD detector.

The minimum-bias trigger efficiency is 100$\%$ for events with D mesons with $\pt > 1~\gevc$ for both pp and p--Pb data sets. For the analyzed data samples, the probability of pile-up from
collisions in the same bunch crossing is below 4$\%$ per triggered pp event and below the percent level per triggered p--Pb event. %
Events in which more than one primary interaction vertex is reconstructed with the SPD detector are rejected, which effectively removes the impact of in-bunch pile-up events on the analysis.
The contribution of particles from pile-up of pp collisions in different bunch crossings is also negligible due to the selections applied to the tracks used in this analysis and the large interval between subsequent bunch crossings in the data samples used. %

\subsection{Monte-Carlo simulations}
\label{sec:MC}
Monte-Carlo simulations including a complete description of the ALICE detector
are used to calculate the corrections for the azimuthal-correlation distributions evaluated from data.
The distribution of the collision vertex along the beam line, the conditions of all the ALICE detectors, and their evolution with time during the pp and p--Pb collision runs are
taken into account in the simulations. Proton-proton collisions are simulated with the $\pythia$~6.4.21 event generator~\cite{pythia}
with the Perugia-0 tune (tune number 320)~\cite{PerugiaTunes} while p--Pb collisions are simulated using the
HIJING v1.36 event generator~\cite{hijing}. For the calculation of D-meson reconstruction efficiencies PYTHIA simulations
of pp collisions are used, requiring that in each
event a $\ccbar$ or $\bbbar$ pair is present. In the simulation used for the analysis of p--Pb data, a p--Pb collision
simulated with HIJING is added on top of the PYTHIA event. The generated particles are transported through
the ALICE apparatus using the GEANT3 package~\cite{geant}.

The measured angular-correlation distributions are compared to simulation results
obtained with the event generators $\pythia$~6.4.25 \cite{pythia} (tunes number 320, 327, and 350, corresponding to the
reference versions of the Perugia-0, Perugia-2010, and Perugia-2011 sets~\cite{PerugiaTunes}, respectively), $\pythia~8.1$ (tune 4C)~\cite{pythia8}, $\powheg$~\cite{powheg1,powheg2} coupled to PYTHIA (Perugia-2011 tune), and EPOS~3.117~\cite{epos,eposbis,epos3} (referred to as EPOS 3 hereafter). PYTHIA simulations utilise LO-pQCD matrix elements for $2\rightarrow 2$ processes, along with
a leading-logarithmic $\pt$-ordered parton shower, the Lund string model for
hadronisation, and an underlying-event simulation including Multiple-Parton Interactions (MPI). %
With respect to older tunes, the Perugia tunes use different initial-state radiation and final-state radiation models. One of the main differences is
that the parton shower algorithm is based on a $\pt$-ordered evolution rather than a virtuality-ordered one. %
Significant differences in the treatment of colour
reconnection, MPI, and the underlying event were also introduced. Perugia 0 is the first of the series. The Perugia-2010 tunes
differ from those of Perugia-0 in the amount of final-state radiation and by a modification of the
high-$z$ fragmentation (inducing a slight hardening of the spectra). They are expected to better reproduce observables related to the jet
shape. The first LHC data, mainly from multiplicity and underlying-event related measurements, were considered for the Perugia-2011 tunes. PYTHIA~8.1
also includes several improvements in the treatment of MPI and colour reconnection~\cite{pythia8}. In the simulations
done with $\sqrt{s}=5.02~\tev$, the centre-of-mass frame is boosted in rapidity by $\Delta y_{\rm{NN}} = 0.465$ in order to reproduce
the rapidity shift of the reference frame of the nucleon-nucleon collision in the p--Pb collision system.

$\powheg$ is a NLO-pQCD generator~\cite{powheg1,powheg2} that, coupled to parton shower programs (e.g.~from PYTHIA or HERWIG~\cite{herwig}),
can provide exclusive final-state particles, maintaining the next-to-leading order
accuracy for inclusive observables. The charm-production cross sections obtained with \mbox{POWHEG+PYTHIA} are consistent with FONLL~\cite{fonll}
and GM-VFNS~\cite{gmvfns} calculations within the respective uncertainties, and are in agreement with measured D-meson production cross sections within
the model and experimental uncertainties~\cite{comparisonFONLLpowheg,gmvfnsPOWHEGtoDATA}. The \mbox{POWHEG+PYTHIA} simulations presented in this paper are
obtained with the POWHEG BOX framework~\cite{powheg3,powhegHF} and the tune Perugia 2011 of PYTHIA 6.4.25.
For the comparison with the measured p--Pb collision data, parton distribution functions (PDFs) corrected
for nuclear effects (CT10nlo~\cite{CT10nlo} with EPS09~\cite{EPS09})
are used. In addition, a boost in rapidity by $\Delta y_{\rm{NN}} = 0.465$ is applied to the partons generated with POWHEG before the PYTHIA parton shower process.

EPOS~3~\cite{epos,eposbis,epos3} is a Monte-Carlo event generator based on a
3+1D viscous hydrodynamical evolution starting from flux tube initial conditions, which
are generated in the Gribov-Regge multiple-scattering framework. Individual scatterings are
referred to as Pomerons, and are identified with parton ladders. Each parton ladder is composed of a
pQCD hard process with initial and final state radiation. Non-linear effects are considered by means of a
saturation scale. The hadronisation is performed with a string fragmentation procedure. Based on these
initial conditions, the hydrodynamical evolution can be applied on the dense core of the collision. An evaluation within the EPOS~3 model shows that the energy density
reached in pp collisions at $\sqrts=7~\tev$ is high enough to apply such hydrodynamic evolution~\cite{epos3}.
\section{Data analysis}
\label{sec:Analysis}
The analysis procedure consists of three main parts, which are described in the following subsections: D-meson reconstruction and selection of primary particles to be used in the correlation analysis (Section~\ref{sec:sigextraction}), construction of azimuthal-correlation distribution and corrections, including the subtraction of combinatorial background and beauty feed-down contributions (Section~\ref{sec:corrections}), extraction of correlation properties via fits to the azimuthal distributions (Section~\ref{sec:fitdescription}).

\subsection{D-meson and associated-particle reconstruction}
\label{DsignalExtraction}
\begin{figure}[!t]
\includegraphics[width=\linewidth]{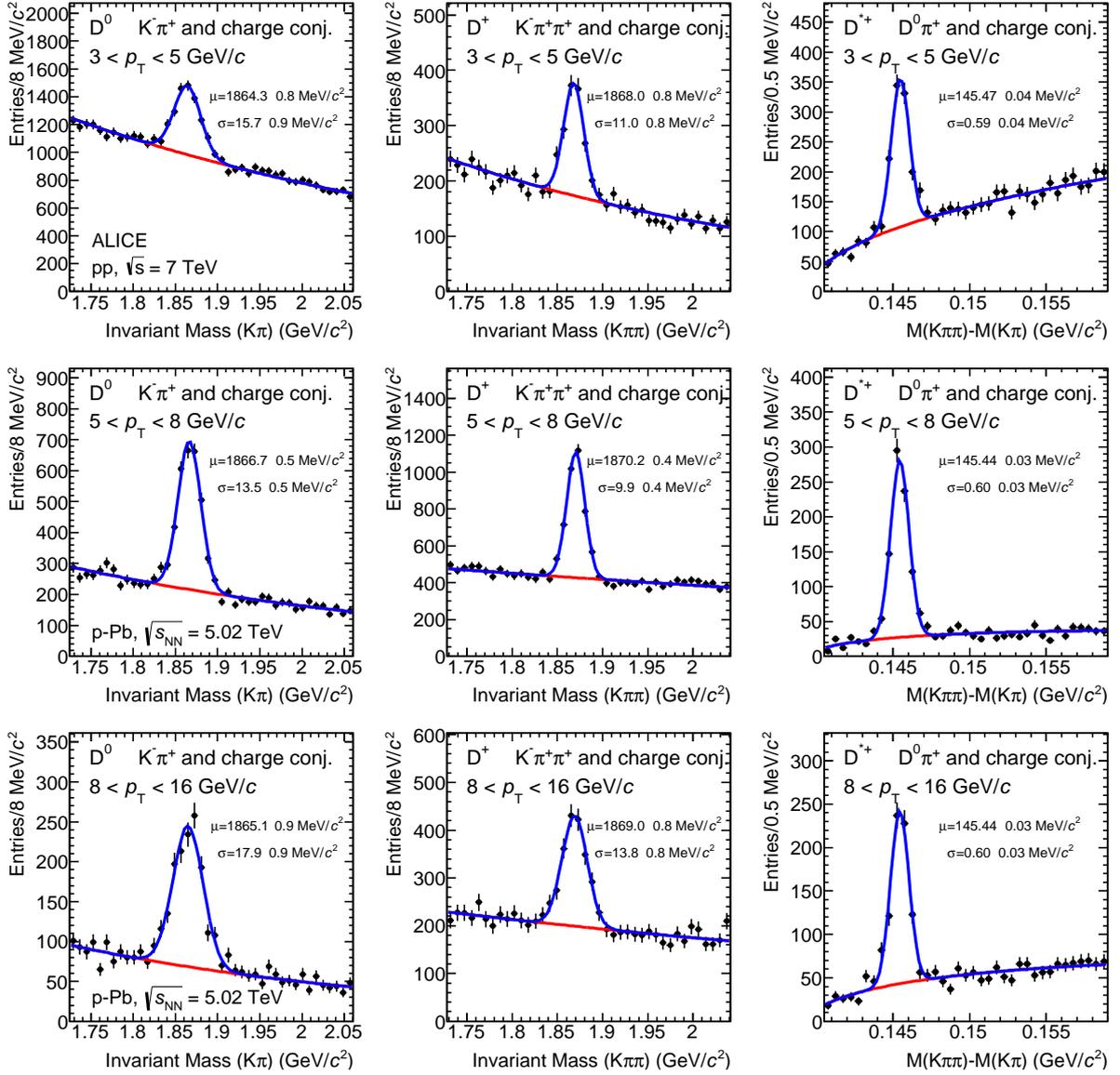}
\caption{Distributions of $\Dzero$ (left column) and $\Dplus$ (middle column) candidates invariant mass and of
the $\Dstar$ candidates invariant-mass difference (right column). The distributions are shown for pp collisions in the $3<\pt<5~\gevc$ range (top row) and for p--Pb collisions in the $5<\pt<8~\gevc$ (middle row) and $8<\pt<16~\gevc$ (bottom row) ranges. The fits
to the invariant mass distributions and the Gaussian mean and sigma values are also shown.}
\label{fig:massPlots}
\end{figure}

The correlation analysis is performed by associating D mesons ($\Dzero$, $\Dplus$, $\Dstar$ mesons and their antiparticles), defined as ``trigger'' particles, with charged primary particles in the same event, and excluding those coming from the decay of the trigger D mesons themselves. 
The $\Dzero$, $\Dplus$, $\Dstar$ mesons and their charge conjugates are reconstructed via their hadronic decay
channels $\DtoKpi$, with Branching Ratio (BR) of (3.88$\pm$0.05)$\%$, $\DtoKpipi$, BR of (9.13$\pm$0.19)$\%$, and $\DstartoDpi$, BR of (67.7$\pm$0.5)$\%$~\cite{PDG}. %
The D-meson signal extraction is based on the reconstruction of decay vertices
displaced from the primary vertex by a few hundred microns and on the identification of the decay-particle species. The
same selection procedures used for
the measurements of D-meson production in pp and p--Pb collisions
at $\sqrts=7~\tev$ and $\sqrtsNN=5.02~\tev$, respectively, are adopted~\cite{DmesonPP,ALICEDmesonpPb}. For both the pp and p--Pb
data sets, $\Dzero$ and $\Dplus$  candidates are formed by combining two or three tracks, respectively, with each track satisfying $|\eta|<0.8$ and
$\pt>0.3~\gevc$. Additionally, $\Dzero$ and $\Dplus$ daughter tracks are required to have at least 70 out of a maximum of 159 possible associated space points in the TPC,
a $\chi^{2}/$NDF of the momentum fit in the TPC smaller than 2, and at least 2 out of 6
associated hits in the ITS. $\Dstar$ candidates are formed combining $\Dzero$ candidates with tracks with one point in the SPD, $\abs{\eta}<0.8$ and $\pt>0.1~\gev/c$. The main variables used to reject the combinatorial background
are the separation between primary and secondary vertices, the distance of closest approach (DCA)
of the decay tracks to the primary vertex, and the angle between the reconstructed D-meson
momentum and the flight line defined by the primary and secondary vertices. A tighter selection is
applied for p--Pb collisions with respect to pp collisions to reduce the larger combinatorial background. Charged kaons and pions are identified using the TPC and TOF detectors. A $\pm 3\sigma$ cut around the expected value
for pions and kaons is applied on both TPC and TOF signals. The D mesons are selected in a fiducial rapidity range varying
from $|y_{\rm lab}|<0.5$ at low $\pt$ to $|y_{\rm lab}|<0.8$ for D mesons with $\pt > 5~\gevc$ in order to avoid cases in which the decay tracks are close to the edge of the detector, where the acceptance decreases steeply. The $\Dzero$ and $\Dplus$ raw yields are
extracted using fits to the distributions of invariant mass $M({\rm K^-}\pi^{\rm +})$ and $M({\rm K^-}\pi^{\rm +}\pi^{\rm +})$, respectively, with a function composed of a Gaussian term for the signal and an exponential term that models the combinatorial background. In the case of the $\Dstar$, the raw yield is obtained by fitting the invariant-mass difference $\Delta M=M({\rm K^-}\pi^{\rm +}\pi^{\rm +}) - M({\rm K^-}\pi^{\rm +})$, using a Gaussian function for the signal and a threshold function multiplied by an exponential ($a\sqrt{\Delta M - M_\pi} \cdot e^{b(\Delta M - M_\pi)}$) to describe the background. Relatively wide D-meson $\pt$ intervals ($3<\pt<5~\gevc$, $5<\pt<8~\gevc$, $8<\pt<16~\gevc$ for pp collisions and $5<\pt<8~\gevc$, $8<\pt<16~\gevc$ for p--Pb collisions) are chosen to reduce the statistical fluctuations in the azimuthal-correlation distributions.
Figure~\ref{fig:massPlots} shows the $\Dzero$ and $\Dplus$ invariant mass, and $\Dstar$ invariant-mass difference distributions in the $3<\pt<5~\gevc$ interval for pp collisions and in the $5<\pt<8~\gevc$, $8<\pt<16~\gevc$ intervals for p--Pb collisions. The fits used to evaluate the raw yields are also shown.

The statistical uncertainty of the D-meson raw yields in the $\pt$ intervals analyzed varies from about 5\% to 8\% (3\% to 5\%) in pp (p--Pb)
collisions for the $\Dzero$ and $\Dplus$ mesons and from about 5\% to 6\% (5\% to 10\%) for the $\Dstar$ mesons, depending on $\pt$. For both
collision systems, the signal over background ratio of the signal peaks is between 0.2 and 1 for the $\Dzero$ and $\Dplus$ mesons, and up
to 2.6 for the $\Dstar$ meson. In the interval $3<\pt<5~\gevc$ the D-meson yield can be extracted from the invariant mass distribution
with statistical uncertainty smaller than 3\% in both pp and p--Pb collisions. However, in the latter case, the near- and away-side peaks of the azimuthal-correlation distribution, that have a small amplitude at low D-meson $\pt$, cannot be disentangled from the statistical fluctuations of the baseline, which is related to the multiplicity of the event and thus higher in p--Pb than in pp collisions. Therefore, for this $\pt$ interval, the results are shown only for pp collisions.

Associated particles are defined as all charged primary particles with $\pt^{\rm assoc}>0.3$ $\gevc$ and with pseudorapidity
$|\eta|<0.8$, except for the decay products of the trigger D meson. Particles coming from other weak decays or
originating from interactions with the detector material are defined as secondary particles and are discarded. %
Reconstructed tracks with at least 70 points in the TPC and 3 in the ITS, and a $\chi^{2}/$NDF of the momentum
fit in the TPC smaller than 2 are associated to D-meson candidates. Using Monte~Carlo simulations (see Section~\ref{sec:MC}), these selection criteria yield an average track reconstruction efficiency for charged primary particles of about $85$\% in the pseudorapidity range $|\eta|<0.8$ and in the interval $0.3<\pt<24~\gevc$, with variations contained within $\approx 5$\% for $\pt<1.5~\gev/c$. Negligible variations are observed at higher $\pt$. The contamination of secondary particles is removed by requiring
the DCA of
the associated tracks to the primary vertex to be less than $2.5~\mm$ in the transverse ($x,y$) plane and less than $1~\cm$ along the beam line
($z$ direction). This selection identifies primary particles with a purity ($p_{\rm prim}$) of approximately $96\%$ and an
efficiency higher than $99\%$, also for particles originating from decays of charm or beauty hadrons, which can be
displaced by several hundred micrometers from the primary vertex. The
purity is independent of $\pt$ in the measured $\pt$ range.
For the $\Dzero$-meson case, the low-$\pt$ pion produced from the $\Dstar\rightarrow\Dzero\pi^+$ decay
is removed from the sample of associated particles by rejecting tracks that yield a $\Delta M$ compatible within $3\sigma$ with the
value expected for $\Dstar$ mesons. It was verified with Monte Carlo simulations
that this selection rejects more than 99\% of the pions from $\Dstar$ decays in all D-meson $\pt$ intervals considered and has
an efficiency larger than $99\%$ for primary particles with $\pt>0.3~\gevc$. %

\label{sec:sigextraction}
\subsection{Azimuthal-correlation distributions and corrections}
\label{corrections}
D-meson candidates with invariant mass ($M$) in the range $|M-\mu|<2\sigma$ (peak region), where $\mu$ and $\sigma$ denote
the mean and width of the Gaussian term of the invariant-mass fit function, are correlated to tracks selected with the criteria described above, and
the difference in the azimuthal angle ($\Delta\varphi$) and in pseudorapidity ($\Delta\eta$) of each pair is computed. In order
to correct for the acceptance and reconstruction efficiency ($\rm {Acc} \times \epsilon$) of the associated tracks and for the variation of ($\rm {Acc} \times \epsilon$)
of prompt D mesons inside a given $\pt$ interval, a weight
equal to the inverse of the product of both ($\rm {Acc} \times \epsilon$) is assigned to each pair. The dependence of the associated-track efficiency on transverse momentum, pseudorapidity, and position of the primary vertex along the beam axis is taken into account. The dependence of the track reconstruction efficiency on the event multiplicity is negligible and therefore neglected. %
The reconstruction efficiency of prompt D mesons is calculated as a function of $\pt$ and event multiplicity. It is on the order of few percent in the lowest D-meson $\pt$ interval, about 20$\%$ at high $\pt$~\cite{DmesonPP,ALICEDmesonpPb}, and varies within each $\pt$ interval by up to a factor 2-3 (1.5-2) at low (high) $\pt$, depending on the D-meson species and collision system. The D-meson ($\rm {Acc} \times \epsilon$) factor also accounts for the $\pt$-dependent fiducial rapidity range of the selected D mesons (Sec.~\ref{DsignalExtraction}) in order to normalise the results to one unit of rapidity.

The obtained distribution, $C(\Delta\varphi,\Delta\eta)_{\rm peak}$, also includes the angular correlation of combinatorial D-meson candidates in the peak range, which is a source of background and needs to be subtracted. %
This contribution is estimated via the per-trigger correlation distribution of background candidates in the sideband invariant-mass range, $1/B_{\rm sidebands} \times C(\Delta\varphi,\Delta\eta)_{\rm sidebands}$, where $B_{\rm sidebands}$ is the amount of background in the sideband region $4\sigma<|M-\mu|<8\sigma$ (right side only, $4\sigma<M-\mu<15\sigma$, in the case of $\Dstar$ mesons).
The term $C(\Delta\varphi,\Delta\eta)_{\rm sidebands}$ represents the correlation distribution obtained as described above, but selecting trigger D-meson candidates with invariant mass in the sidebands. The background contribution is then subtracted from $C(\Delta\varphi,\Delta\eta)_{\rm peak}$ after  being normalised to the amount of combinatorial background in the peak region, $B_{\rm peak}$. The latter is obtained from the counts in the invariant-mass distribution in the peak region, after subtracting the signal, $S_{\rm peak}$, estimated from the invariant-mass fit. Note that $S_{\rm peak}$, $B_{\rm peak}$ and $B_{\rm sidebands}$ are calculated from the invariant-mass distributions weighted by the inverse of the prompt D-meson reconstruction efficiency.

The correlation distributions $C(\Delta\varphi,\Delta\eta)_{\rm peak}$ and $C(\Delta\varphi,\Delta\eta)_{\rm sidebands}$ are corrected for the limited detector acceptance
and spatial inhomogeneities using the event mixing technique. In this approach, D-meson candidates found in a given event are correlated with charged
tracks from other events with similar multiplicity and primary-vertex position along the beam axis.
The distribution obtained from the mixed events, ${\rm ME}(\Delta\varphi,\Delta\eta)$, shows a typical triangular shape as a function of $\Delta\eta$, due to the limited $\eta$ coverage of the detector, and is approximately flat as a function of $\Delta\varphi$. The event-mixing distribution is rescaled by its average value in the range ($-0.2<\Delta\varphi<0.2$,$-0.2<\Delta\eta<0.2$) and its inverse is used as a map to weight the distributions $C(\Delta\varphi,\Delta\eta)_{\rm peak}$ and $C(\Delta\varphi,\Delta\eta)_{\rm sidebands}$. A correction for the purity of the primary-particle
sample ($p_{\rm prim}$, see Sec.~\ref{DsignalExtraction}) is applied and the per-trigger normalisation
is obtained dividing by $S_{\rm peak}$. The above procedure is summarised in Equation~\ref{eqCinclusiveDef}, where the notation $\tilde{C}$ refers to angular-correlation distributions normalised by the number of trigger particles: %
\begin{linenomath}
  \begin{flalign}
    \tilde{C}_{\rm inclusive}(\Delta\varphi,\Delta\eta) &= \frac{p_{\rm prim}}{S_{\rm peak}}\left(\left.\frac{C(\Delta\varphi,\Delta\eta)}{{\rm ME}{(\Delta\varphi,\Delta\eta)}}\right|_{\rm peak}
      -\frac{B_{\rm peak}}{B_{\rm sidebands}}\left.\frac{C(\Delta\varphi,\Delta\eta)}{{\rm ME}{(\Delta\varphi,\Delta\eta)}}\right|_{\rm sidebands}\right), &     \label{eqCinclusiveDef} \\
    {\rm ME}(\Delta\varphi,\Delta\eta)&=\left(\frac{C(\Delta\varphi,\Delta\eta)}{\langle C(\Delta\varphi,\Delta\eta)\rangle_{|\Delta\varphi|,|\Delta\eta|<0.2}}\right)_{\rm Mixed~Events} . & \nonumber
  \end{flalign}
\end{linenomath}

Finally, the per-trigger azimuthal
distribution $\tilde{C}_{\rm inclusive}(\Delta\varphi)$ is obtained by integrating $\tilde{C}_{\rm inclusive}(\Delta\varphi,\Delta\eta)$
in the range $\abs{\Delta\eta}<1$.

It was verified using Monte-Carlo simulations based on PYTHIA (Perugia-2011 tune) that the per-trigger azimuthal correlation of D mesons and
secondary particles not rejected by the track selection has a $\Delta\varphi$-dependent modulation with a maximum variation of $7\%$ with respect to the azimuthal correlation
of D mesons and primary particles. This $\Delta\varphi$-dependent contamination has a negligible impact on the final results, considering the $4\%$ level
of contamination of secondary particles in the sample of associated tracks, hence, it was neglected.

A fraction of the reconstructed D mesons consists of secondary
D mesons coming from B-meson decays. The topological
cuts, applied to reject combinatorial background, preferentially select displaced vertices, yielding
a larger (by about a factor 2 for $\Dzero$ mesons in the measured $\pt$ range) efficiency for secondary
D mesons than for prompt D mesons. Therefore, the fraction $f_{\rm prompt}$ of reconstructed prompt D mesons
does not coincide with the natural fraction and depends on the analysis details. The different fragmentation, as well
as the contribution of B-meson decay particles and a possible different contribution of gluon splitting to charm- and beauty-quark production, imply a different angular-correlation distribution of prompt and secondary
D mesons with charged particles, as it was verified with the Monte-Carlo simulations described in Section~\ref{sec:MC}. The
contribution of feed-down D mesons to the measured angular correlation is subtracted as follows:
\begin{equation}
  \tilde{C}_{\rm prompt}(\Delta\varphi)=\frac{1}{f_{\rm prompt}}\left(\tilde{C}_{\rm inclusive}(\Delta\varphi)-(1-f_{\rm prompt})\tilde{C}_{\rm \operatorname{feed-down}}^{\rm MC~templ}(\Delta\varphi) \right) .
  \label{eqFeedDown}
\end{equation}
In Equation~\ref{eqFeedDown}, $\tilde{C}_{\rm prompt}(\Delta\varphi)$ is
the per-trigger azimuthal-correlation distribution after the subtraction of the feed-down contribution, $f_{\rm prompt}$ is the fraction of
prompt D mesons and $\tilde{C}_{\rm feed-down}^{\rm MC~templ}(\Delta\varphi)$ is a template for the azimuthal-correlation distribution of
the feed-down component. Using the same method described in~\cite{DmesonPP}, $f_{\rm prompt}$ was evaluated
on the basis of FONLL calculations of charm and beauty $\pt$-differential production cross sections~\cite{fonll}
and of the reconstruction efficiencies of prompt and secondary D mesons, calculated using Monte-Carlo simulations. The value of $f_{\rm prompt}$, which
depends on the D-meson species and varies as a function of $\pt$, is estimated to be larger than $75\%$. %
The azimuthal correlation of feed-down D mesons, $\tilde{C}_{\rm feed-down}^{\rm MC~templ}$, was obtained from PYTHIA (tune Perugia 2011~\cite{PerugiaTunes})
simulations of pp collisions at $\sqrts=7~\tev$ and $\sqrts=5.02~\tev$ for the analysis of pp and p-Pb data, respectively. In order
to avoid biases related to the different event multiplicity in real and simulated events, the correlation distribution
was shifted to have its minimum coinciding with the baseline of the data azimuthal-correlation distribution before
feed-down subtraction.
A difference smaller than $8\%$ was observed in the simulation between the baseline values of the azimuthal-correlation
distributions for prompt and feed-down D mesons. Considering the typical values of $f_{\rm prompt}$, this difference
results in a shift of the baseline of $\tilde{C}_{\rm prompt}(\Delta\varphi)$ smaller than $2\%$, negligible with respect to the other
uncertainties affecting the measurement.
\label{sec:corrections}
\subsection{Characterization of azimuthal-correlation distributions}
\label{sec:fitdescription}
In order to quantify the properties of the measured azimuthal correlations, the following fit function is used:
\begin{equation}
f(\Delta\varphi)=b+\frac{A_{{\rm NS}}}{\sqrt{2\pi}\sigma_{\rm fit,NS}}e^{-\frac{(\Delta\varphi)^{2}}{2\sigma^{2}_{{\rm fit,NS}}}}+\frac{A_{{\rm AS}}}{\sqrt{2\pi}\sigma_{{\rm fit,AS}}}e^{-\frac{(\Delta\varphi-\pi)^{2}}{2\sigma^{2}_{{\rm fit,AS}}}}.
\label{eq:fitfunction}
\end{equation}
 It is composed of two Gaussian terms describing the near- and away-side
peaks and a constant term describing the baseline. A periodicity condition is also imposed to the function, requiring $f(0) = f(2\pi)$.

The integrals of the Gaussian terms, $A_{{\rm NS}}$ and $A_{{\rm AS}}$, correspond to the associated-particle yields
for the near (NS)- and away (AS)-side peaks, respectively, while $\sigma_{{\rm fit,NS}}$ and $\sigma_{{\rm fit,AS}}$ quantify the widths of the
correlation peaks. By symmetry considerations, the mean of the Gaussian functions are fixed to $\Delta\varphi = 0$ and
$\Delta\varphi = \pi$. The baseline $b$ represents the physical minimum of the $\dphi$ distribution. To limit the effect of statistical fluctuations on the estimate of the associated yields, $b$ is fixed to the weighted average of the points in the transverse region, defined as $\pi/4 < |\Delta\varphi| < \pi/2$, using the inverse of the square of the point statistical uncertainty as weights.
Given the symmetry of the correlation distributions around $\Delta\varphi=0$ and $\Delta\varphi=\pi$, the azimuthal distributions are reported
in the range $0<\Delta\varphi<\pi$ to reduce statistical fluctuations. The effect of a $\vtwo$-like modulation in the $\Delta\varphi$ distribution, which could be present in p--Pb collisions, was estimated and assessed in Section~\ref{sec:Results}.

In the case of the simulations, for which statistical fluctuations
are negligible, the baseline is estimated as the minimum of the azimuthal-correlation distribution. An alternative fitting procedure based on a convolution of two Gaussian functions for the description of the NS peak was performed for Monte Carlo simulations. The resulting NS yields were found to be compatible with those obtained with the standard procedure, with a maximum variation of 7\% (10\%) in pp (p--Pb) collisions in case of EPOS~3 simulations. 

\section{Systematic uncertainties} 
\label{sec:systematics}
The fit of the D-meson invariant-mass distribution introduces
systematic uncertainties on $S_{\rm peak}$ and $B_{\rm peak}$ (Section~\ref{corrections}, Equation~\ref{eqCinclusiveDef}). The uncertainty
on the correlation distribution was estimated by calculating $B_{\rm peak}$ from the integral of the background term
of the invariant-mass fit function in the range $|M-\mu|<2\sigma$ and by varying the fit procedure. In particular, the fit was repeated modeling the background distribution with a linear function and a parabola instead of
an exponential function (for $\Dzero$ and $\Dplus$ mesons only), considering a different histogram binning, and varying the fit
range. %
A $10\%$ systematic uncertainty was estimated from the corresponding variation of the azimuthal-correlation distribution. No significant
trend was observed as a function of $\dphi$ and the same uncertainty was estimated for all D-meson species in all $\pt$-intervals and in both pp and p--Pb collision systems.

A 5\% uncertainty (10\% for $\Dplus$ mesons in p--Pb collisions) arises from the possible dependence of the shape of $\tilde{C}(\Delta\varphi,\Delta\eta)_{\rm sidebands}$ on the sideband range. This source of uncertainty was determined by restricting the invariant-mass sideband window to the intervals $4\sigma<\abs{M-\mu}<6\sigma$ or to $6\sigma<\abs{M-\mu}<8\sigma$ for all the D mesons, and also by considering, for $\Dzero$ and $\Dplus$ mesons, only the left or only the right sideband. 

The uncertainty on the correction for the associated-particle reconstruction efficiency was assessed by varying the
selection criteria applied to the reconstructed tracks, removing the request of at least three associated clusters in the ITS, or
demanding a hit on at least one of the two SPD layers. %
A $\pm 4\%$ uncertainty was estimated for $\pPb$ collisions, while a $^{+10\%}_{-5\%}$ contribution was obtained for the pp analysis, with the +10\% contribution
arising from the request of hits in the SPD. No significant trend in $\dphi$ was observed.%

The uncertainty on the residual contamination from secondary tracks was evaluated by repeating the analysis varying
the cut on the DCA in the $(x,y)$ plane from $0.1~\cm$ to $1~\cm$, and re-evaluating the purity of charged primary particles for each
variation. This resulted in a 5\% (3.5\%) systematic uncertainty in pp ($\pPb$) collisions, independent of $\dphi$ and $\ptAssoc$.

A 5\% systematic effect originating from the correction of the D-meson reconstruction
efficiency was evaluated by applying tighter and looser topological selections on the D-meson candidates. No significant dependence on $\dphi$
was observed and the same uncertainty was estimated for the three D-meson $\pt$ intervals, apart from $\Dplus$ meson in $\pPb$ collisions, for which a 10\% uncertainty was assigned.

The uncertainty on the subtraction of the beauty feed-down contribution was quantified by generating the templates
of feed-down azimuthal-correlation distributions, $\tilde{C}_{\rm feed-down}^{\rm MC~templ}(\Delta\varphi)$ in Equation~\ref{eqFeedDown}, with different
PYTHIA 6 tunes (Perugia 0, Perugia 2010, see Section~\ref{sec:MC}), and by considering the range of $f_{\rm{prompt}}$ values obtained by
varying the prompt and feed-down D-meson $\pt$-differential production cross sections within FONLL uncertainty band, as described
in~\cite{DmesonPP}. The effect on the azimuthal-correlation distributions is $\dphi$ dependent and contained within $8\%$ and is more
pronounced in the near side, in particular in the low and mid D-meson $\pt$ intervals.

The consistency of the whole correction procedure, prior to the feed-down subtraction, was verified
by performing the analysis on simulated events (``Monte-Carlo closure test'') separately for prompt and feed-down D mesons. For prompt
D mesons, no effect was found for both pp and $\pPb$ collision systems. Conversely, for
feed-down D mesons, an overestimate by about 20\% in the near side was found for both collision systems. It was verified that the
source of this excess is related to a bias induced by the topological selection applied to D mesons, that tends to favour cases
with a small angular opening between the products of the beauty-hadron decay, thus between the D meson and the other decay particles. This
effect results in a $\dphi$-dependent overestimate of the feed-down subtracted correlation distribution in the
near side, contained within 2\%.

The systematic uncertainties affecting the $\dphi$-correlation distributions are summarised
in Table~\ref{tab:systUnc} for both pp and $\pPb$ collision systems. The $\dphi$-dependent parts of the uncertainties arising from the feed-down subtraction
and the Monte-Carlo closure test define the $\dphi$-uncorrelated systematic uncertainties. All the other contributions, correlated
in $\dphi$, act as a scale uncertainty. No significant dependence on the transverse momentum of D mesons and associated particles was observed
for both $\dphi$-correlated and uncorrelated uncertainties, except for the feed-down systematic uncertainty.

\begin{table}[ht]
\centering
\begin{tabular}{c|c|c}
\hline\hline
\rule{0pt}{3ex} System         &      pp                     & $\pPb$                     \\
\rule{0pt}{3ex} D-meson species  & $\Dzero, \Dstar, \Dplus$    & $\Dzero, \Dstar$ ($\Dplus$) \\
[0.5ex]
\hline
\rule{0pt}{3ex}
Signal, background normalisation        & $\pm 10\%$    & $\pm 10\%$ \\
Background $\dphi$ distribution  & $\pm 5\%$     & $\pm 5\%$ ($\pm 10\%$) \\
Associated-track reconstruction efficiency     & $+10\%, -5\%$ & $\pm 4\%$ \\
Primary-particle purity & $\pm 5\%$     & $\pm 3.5\%$ \\
D-meson efficiency      & $\pm 5\%$     & $\pm 5\%$ ($\pm 10\%$) \\
Feed-down subtraction   & up to 8\%, $\dphi$ dependent & up to 8\%, $\dphi$ dependent \\
MC closure test         & $-2\%$ (near side) & $-2\%$ (near side), $\pm 2\%$ \\ [0.7ex]
\hline
\end{tabular}
\caption{List of systematic uncertainties for the $\dphi$-correlation distributions in pp and $\pPb$ collisions. See text for details.}
\label{tab:systUnc}
\end{table}

Different approaches were applied to estimate the systematic uncertainty on the near-side peak associated yield
and peak width and on the baseline, obtained from the $A_{\rm NS}$, $\sigma_{{\rm fit,NS}}$, and $b$ parameters
of the fit of the azimuthal-correlation distribution, as described in Section~\ref{sec:fitdescription}. The main source of uncertainty originates from
the definition of the baseline itself, which is connected to the assumption that the observed variation of
the azimuthal-correlation distribution in the transverse region is determined mainly by statistical fluctuations rather
than by the true physical trend. The variation of $A_{\rm NS}$, $\sigma_{{\rm fit,NS}}$, and $b$ values obtained when considering
a $\pm\pi/4$ variation of the $\dphi$ range defining the transverse region is interpreted as the
systematic uncertainty due to the baseline definition. %
In addition, the fits were repeated by moving upwards and downwards the data points by the corresponding value of the $\dphi$-uncorrelated
systematic uncertainty. The final systematic uncertainty
was calculated by summing in quadrature the aforementioned contributions and, for
the associated yields and baseline, also the systematic uncertainty correlated in $\Delta\varphi$. %
The values of the total systematic uncertainties on the near-side peak yield, width, and baseline are reported in Table~\ref{tab:systUnc2}, for two
intervals of transverse momentum of D mesons and associated particles. 
Considering all the measured kinematic ranges, the uncertainties vary from $\pm 12\%$ to $\pm 25\%$ for the near-side peak yield, from $\pm 2\%$ to $\pm 13\%$ for the near-side peak width and from $\pm 11\%$ to $\pm 16\%$ for the baseline. Typically, lower uncertainties are obtained for p-Pb collisions, where the larger available statistics of the correlation distributions allow for a more precise estimate of the baseline height, which constitutes the main source of uncertainty also on the evaluation of the near-side peak associated yield and width. %
\begin{table}[!bh]
\centering
\begin{tabular}{c|c|c|c|c}
\hline\hline
 \small{System}          &        \multicolumn {2}{c|}{pp}                     &     \multicolumn {2}{c} {$\pPb$}                     \\
 \rule{0pt}{4ex}  \multirow{2}{*}{Kinematic range}  &
 \footnotesize{$5<\ptD<8$ $\gevc$,}  &
     \footnotesize{$8<\ptD<16$ $\gevc$,}  &
      \footnotesize{$5<\ptD<8$ $\gevc$,}  &
      \footnotesize{$8<\ptD<16$ $\gevc$,} \\

       & \footnotesize{$0.3<\ptAssoc<1$ $\gevc$} &
        \footnotesize{$\ptAssoc>1$ $\gevc$} &
        \footnotesize{$0.3<\ptAssoc<1$ $\gevc$} &
        \footnotesize{$\ptAssoc>1$ $\gevc$} \\
\hline
\rule{0pt}{3ex}
\small{NS yield}       & $\pm 22\%$    & $\pm 15\%$ & $\pm 17\%$ &$\pm 12\%$ \\
\small{NS width}   & $\pm 10\%$    & $\pm 5\%$ & $\pm 3\%$ &$\pm 3\%$ \\
\small{Baseline}   &  $\pm 13\%$    & $\pm 15\%$ & $\pm 12\%$ &$\pm 11\%$ \\
\hline
\end{tabular}
\caption{List of systematic uncertainties for near-side (NS) peak associated yield, near-side peak width, and baseline in pp and $\pPb$ collisions, for two different kinematic ranges of D mesons and associated particles. See text for details.}\label{tab:systUnc2}
\end{table}

\section{Results}
\label{sec:Results}
\begin{figure}[!t]
\includegraphics[width=\linewidth]{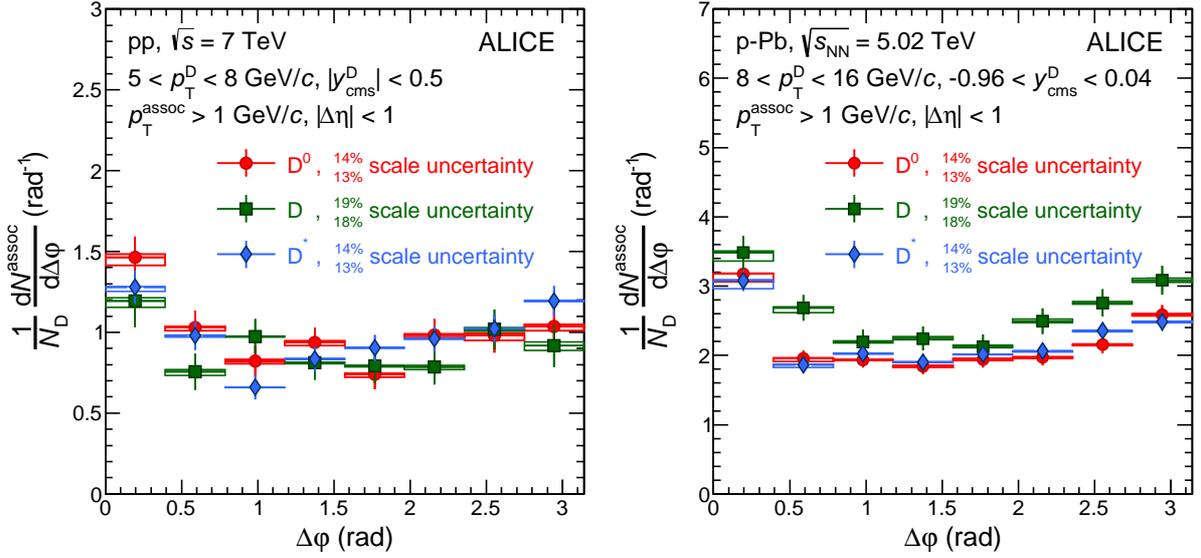}
\caption{Comparison of the azimuthal-correlation distributions of D mesons with charged particles obtained for $\Dzero$, $\Dplus$ and $\Dstar$ mesons for $5<\ptD<8~\gevc$, $\ptAssoc > 1~\gevc$ in pp collisions at $\sqrts=7~\tev$ (left panel) and for $8<\ptD<16~\gevc$, $\ptAssoc > 1~\gevc$
  in $\pPb$ collisions at $\sqrtsNN=5.02~\tev$ (right panel). The statistical uncertainties are shown as error bars, the $\dphi$-uncorrelated systematic uncertainties   as boxes, while the part of systematic uncertainty correlated in $\dphi$ is reported as text (scale uncertainty). The latter is largely uncorrelated   among the D-meson species.}
\label{fig:3mesonComparison}
\end{figure}
\begin{figure}[!t]
\includegraphics[width=\linewidth]{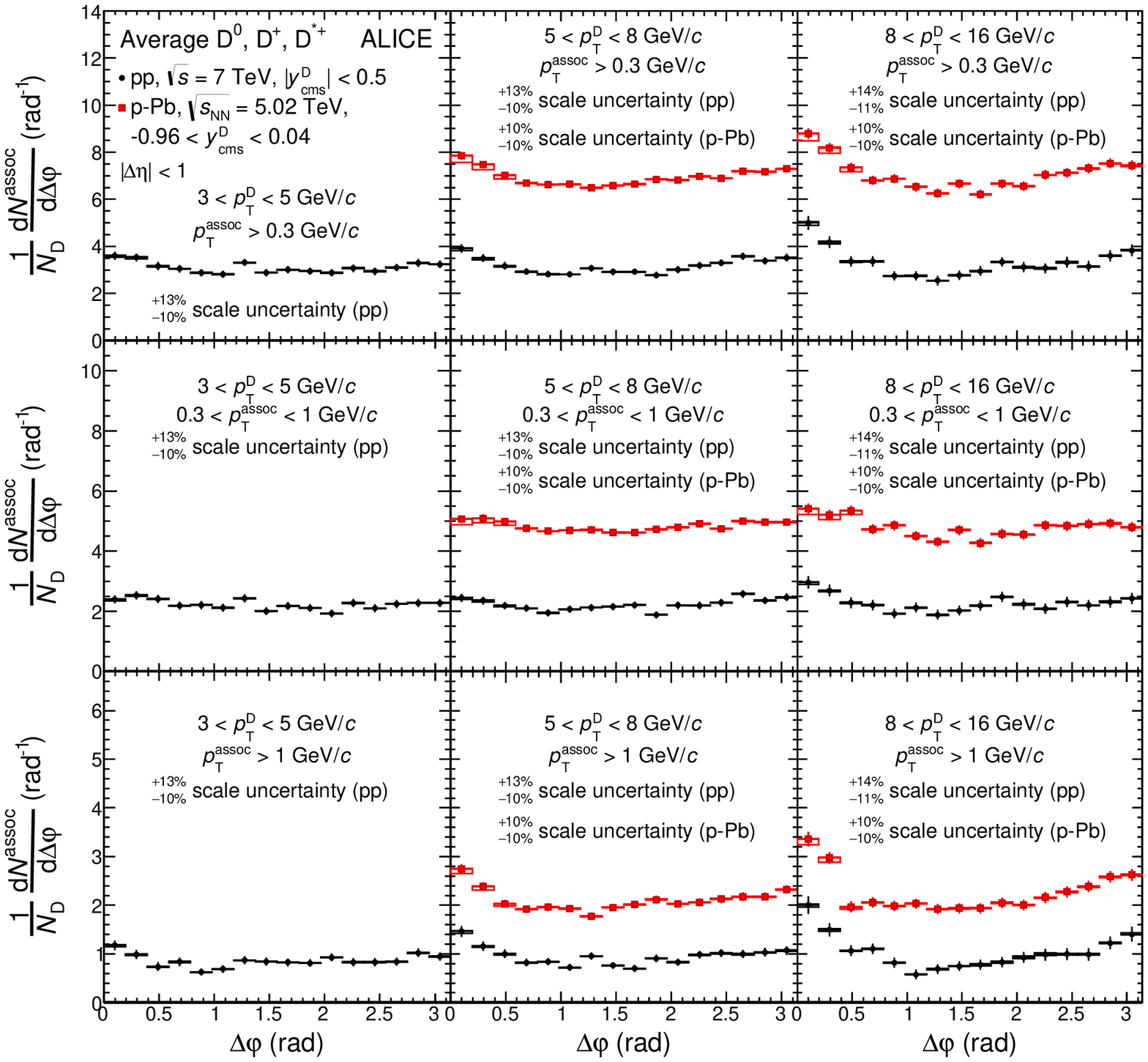}
\caption{Average of the azimuthal-correlation distributions of $\Dzero$, $\Dplus$ and $\Dstar$ mesons
  with $3<\ptD<5~\gevc$ (left column), $5<\ptD<8~\gevc$ (middle column), and $8<\ptD<16~\gevc$ (right column), with charged particles with $\ptAssoc>0.3~\gevc$ (top row), $0.3<\ptAssoc<1~\gevc$ (middle row), and $\ptAssoc>1~\gevc$ (bottom row),
  measured in pp collisions at $\sqrts=7~\tev$ and in p--Pb collisions at $\sqrtsNN=5.02~\tev$. The statistical
  uncertainties are shown as error bars, the $\dphi$-uncorrelated systematic uncertainties
  as boxes, while the part of systematic uncertainty correlated in $\dphi$ is reported as text (scale uncertainty).}
\label{fig:dPhi_Avg_pppPb}
\end{figure}
\begin{figure}[!t]
\centering
\begin{minipage}{\linewidth}
  \centering
\includegraphics[height=0.75\textheight]{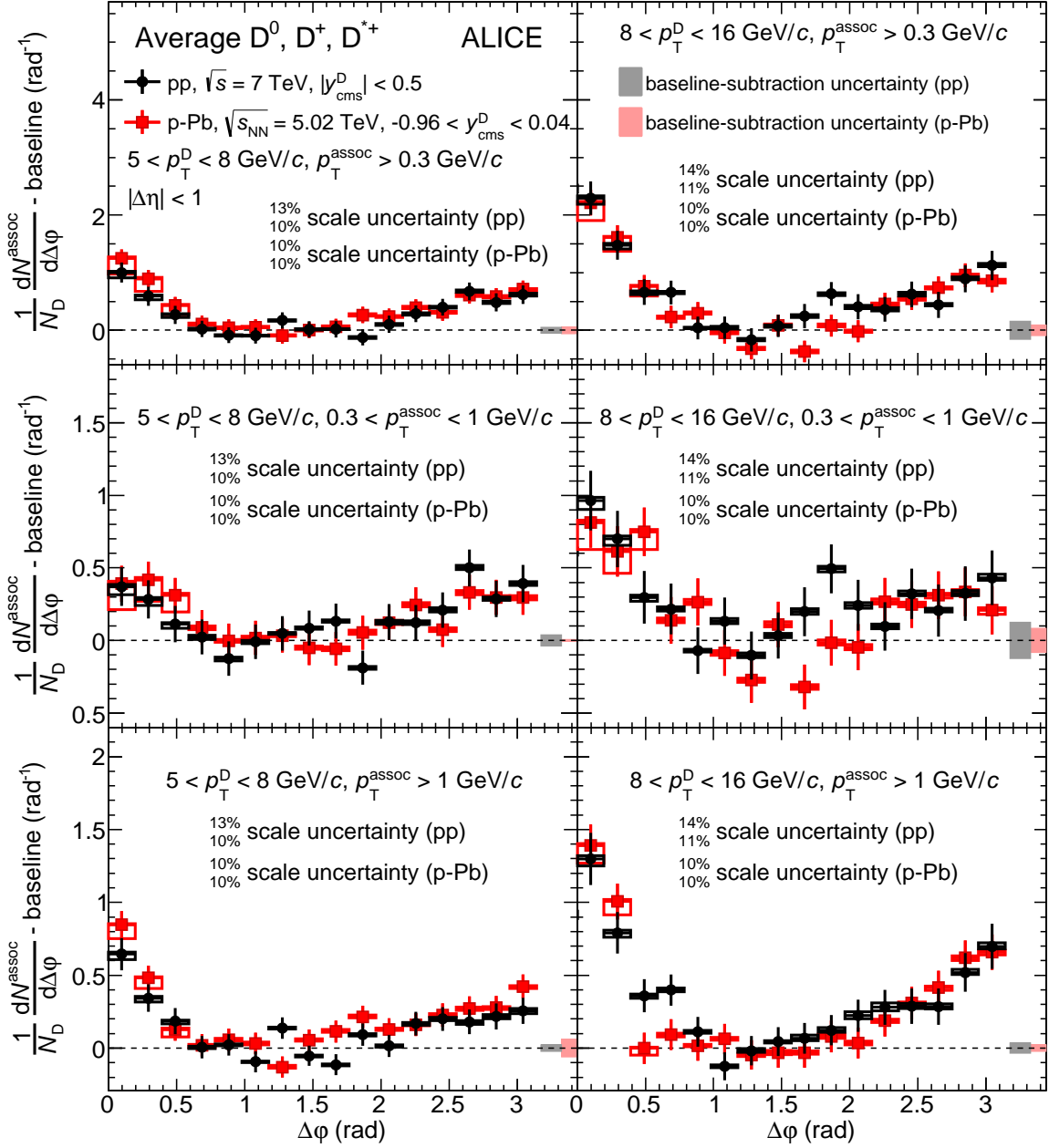}
\end{minipage}
\caption{Comparison of the azimuthal-correlation distributions of D mesons with
$5<\ptD<8~\gevc$ (left column) and $8<\ptD<16~\gevc$ (right column) with charged particles with $\ptAssoc>0.3~\gevc$ (top row), $0.3<\ptAssoc<1~\gevc$ (middle row), and $\ptAssoc>1~\gevc$ (bottom row)
  in pp collisions at $\sqrts=7~\tev$ and in p--Pb collisions at $\sqrtsNN=5.02~\tev$, after baseline subtraction. The statistical
  uncertainties are shown as error bars, the $\dphi$-uncorrelated systematic uncertainties
  as boxes around the data points, the part of systematic uncertainty correlated in $\dphi$ is reported as text (scale uncertainty), the uncertainties
  deriving from the subtraction of the baselines are represented by the boxes at $\dphi>\pi$.}
\label{fig:pp-pPb_Dphi_Compar}
\end{figure}
\begin{figure}[!t]
\centering
  $\vcenter{\hbox{\includegraphics[width=.48\linewidth]{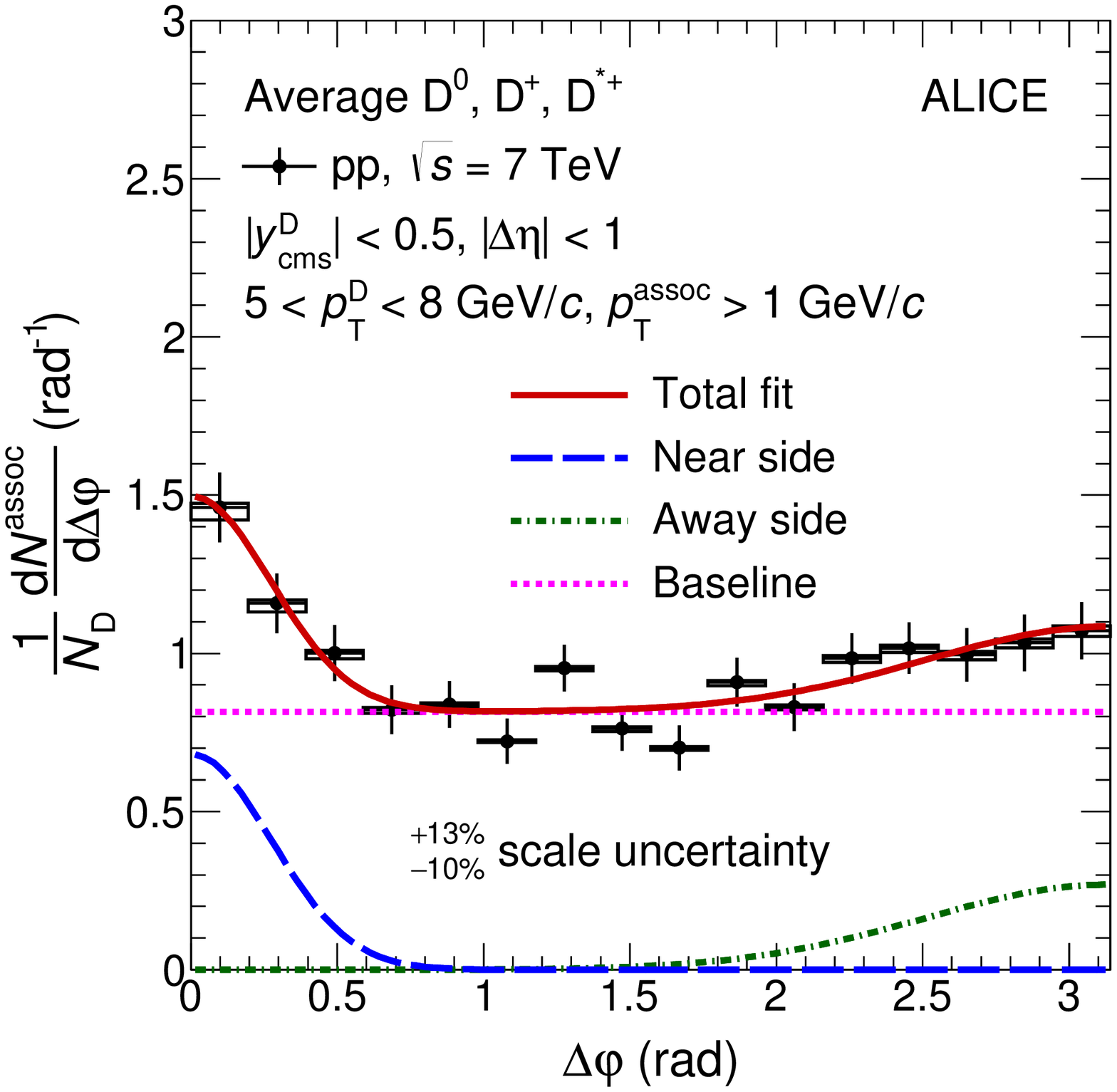}}}$\hspace{0.2cm}
  $\vcenter{\hbox{\includegraphics[width=.48\linewidth]{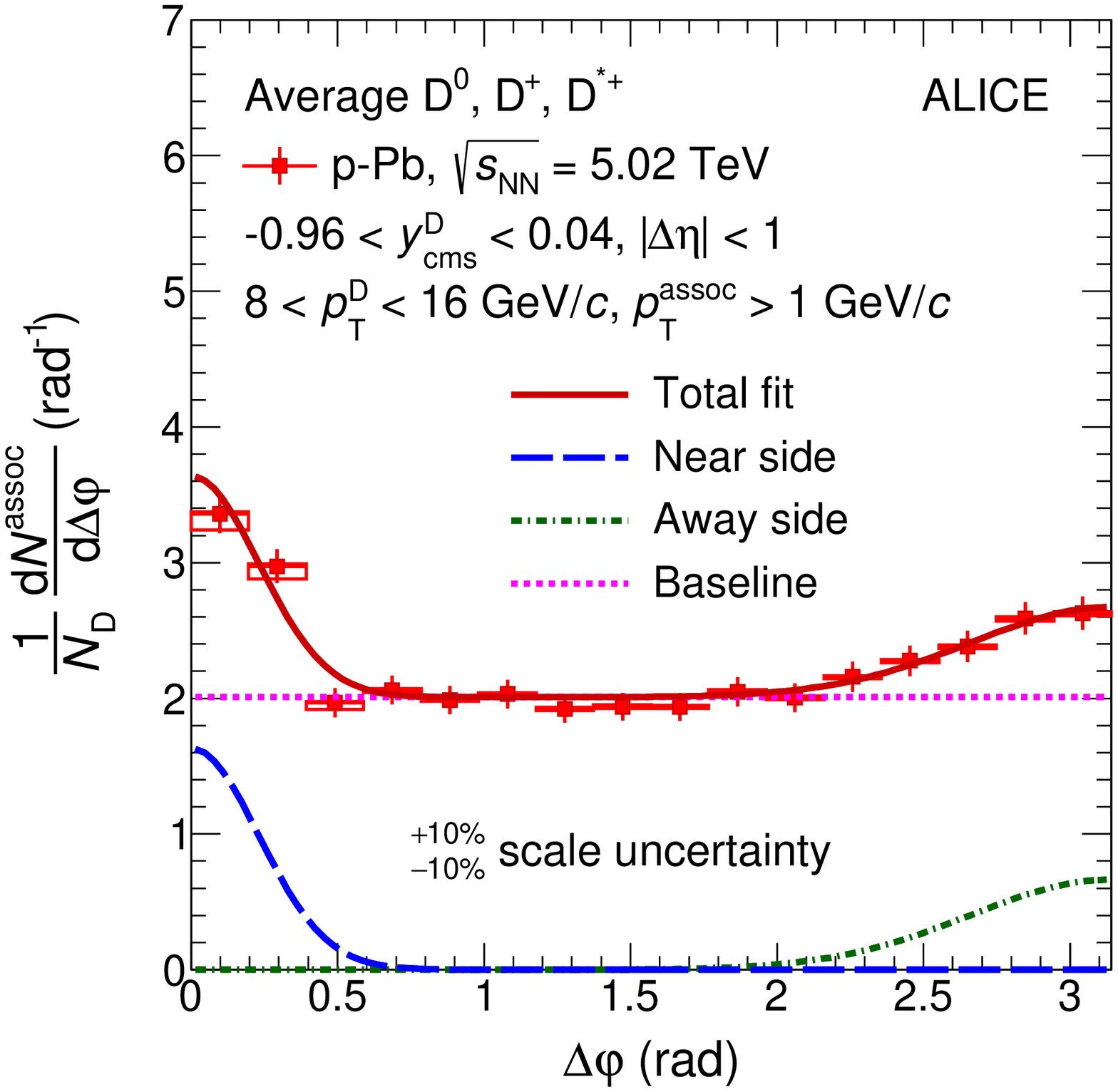}}}$
\caption{Examples of the fit to the azimuthal-correlation distribution, for D mesons with $5<\ptD<8~\gevc$ with charged particles with $\ptAssoc>1~\gevc$ in pp collisions at $\sqrts=7~\tev$ (left), and for D mesons with $8<\ptD<16~\gevc$ with charged particles with $\ptAssoc>1~\gevc$ in p--Pb collisions at $\sqrtsNN=5.02~\tev$ (right). The statistical uncertainties are shown as error bars, the $\dphi$-uncorrelated systematic uncertainties as boxes, while the part of systematic uncertainty correlated in $\dphi$ is reported as text (scale uncertainty). The terms of the fit function described in Section~\ref{sec:fitdescription} are also shown separately: near-side Gaussian function (blue dashed line), away-side Gaussian function (green dashed-dotted line) and baseline constant term (magenta dotted line).}
\label{fig:Fittingazimuth}
\end{figure}
\begin{figure}[!t]
  \centering
  \begin{minipage}{\linewidth}
    \centering
    \includegraphics[width=\linewidth]{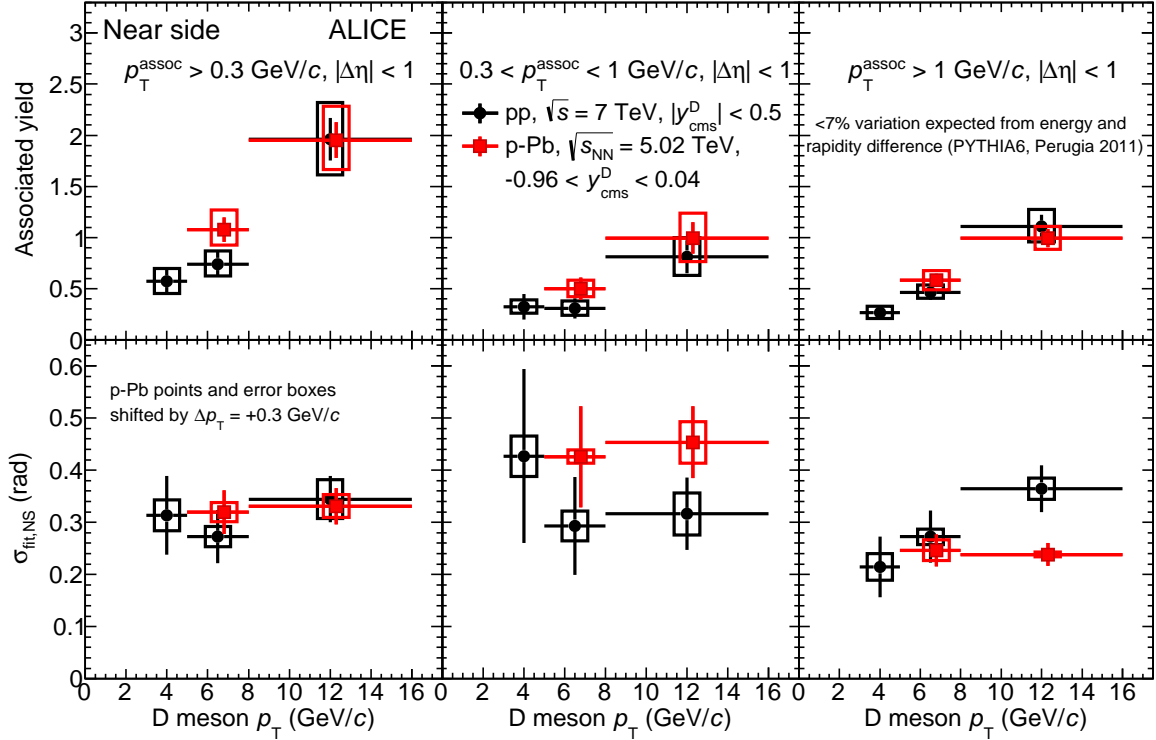}
  \end{minipage}
  \caption{Comparison of the near-side peak associated yield (top row) and peak width (bottom row) in pp and $\pPb$ collisions
    as a function of $\ptD$, for $\ptAssoc>0.3~\gevc$ (left column), $0.3<\ptAssoc<1~\gevc$ (middle column),
    and $\ptAssoc>1~\gevc$ (right column). The points and error boxes for p-Pb collisions are shifted by $\Delta\pt = +0.3~\gevc$. Statistical and systematic uncertainties are shown as error bars and boxes, respectively.}
  \label{fig:ppTopPbFitResComp}
\end{figure}
\begin{figure}[!t]
  \centering
  \resizebox{0.98\textwidth}{!}{\includegraphics{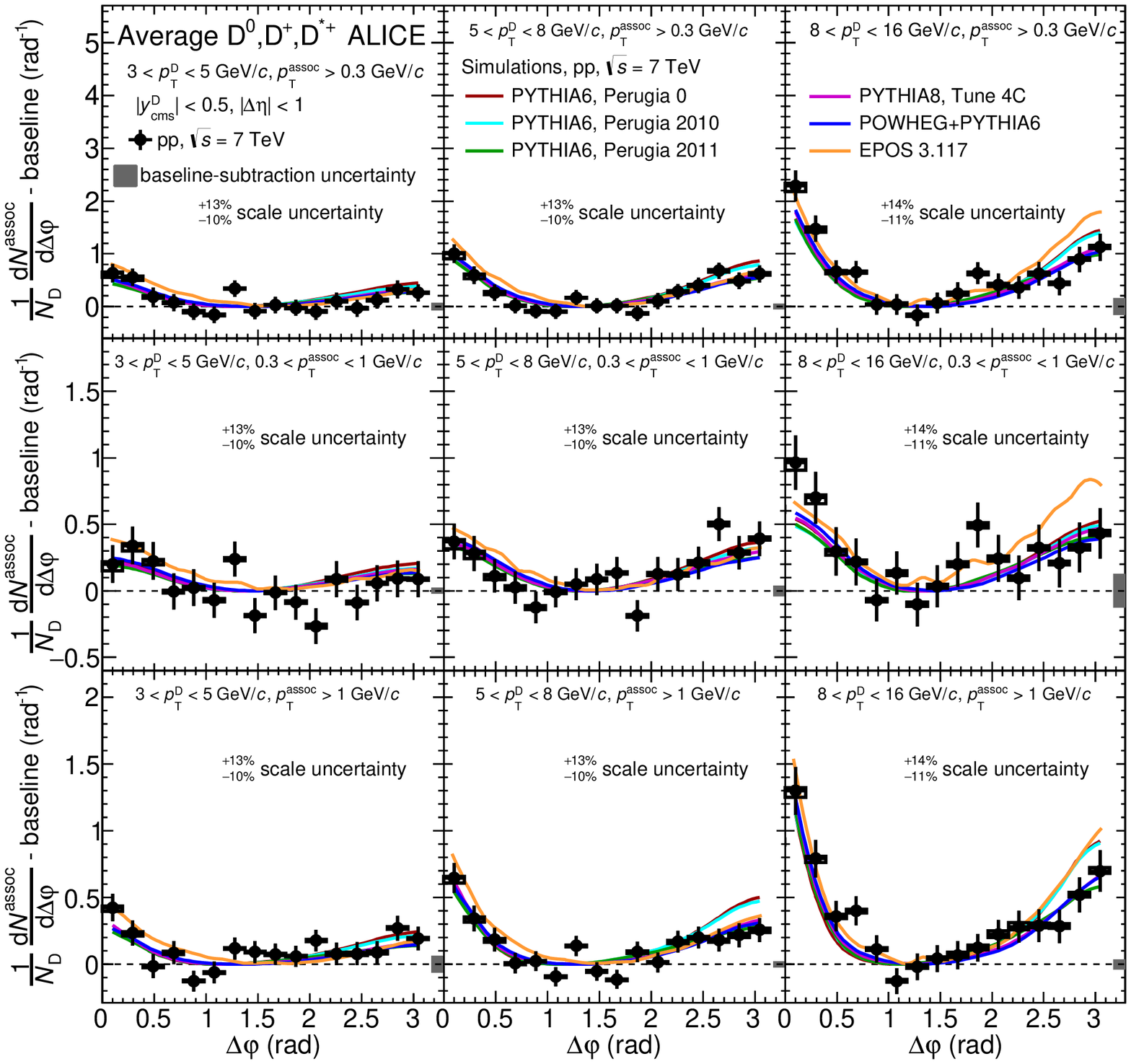}}
  \caption{Comparison of $\dphi$-correlation distributions of D mesons with charged particles measured
    in pp collisions at $\sqrts=7~\tev$ and Monte-Carlo simulations performed with different event generators, after
    the subtraction of the baseline. The statistical and systematic uncertainties of the measured distributions
    are displayed as in Figure~\ref{fig:pp-pPb_Dphi_Compar}.}%
  \label{fig:PYTHIA_Comp}
\end{figure}
The azimuthal-correlation distributions of $\Dzero$, $\Dplus$, $\Dstar$ mesons with
charged particles with $\ptAssoc>1~\gevc$ are compared in Figure~\ref{fig:3mesonComparison} for $5<\ptD<8~\gevc$ in pp collisions (left panel)
and for $8<\ptD<16~\gevc$ in $\pPb$ collisions (right panel). The distributions obtained with the three D-meson species are
compatible within the quadratic sum ($w_{i}$, $i=\Dzero$, $\Dplus$, $\Dstar$) of
the statistical uncertainty and of the systematic uncertainties on the signal, background normalisation, and on the background shape
(see Table~\ref{tab:systUnc}), that are uncorrelated among the three meson species. The
$\Dzero$-, $\Dplus$-, $\Dstar$-meson data are averaged using $1/w_{i}^{2}$ as weights. %
The averages of the distributions are shown, for all the considered kinematic ranges, in
Figure~\ref{fig:dPhi_Avg_pppPb} for pp and $\pPb$ collisions. A rising trend of the height of the near-side peak with increasing
D-meson $\pt$ is observed for both collision systems. A similar trend is present for hadron-hadron correlations measured at Tevatron and LHC energies~\cite{CDFhhJet,CMSunderlyingEvent,ATLAShhpp,ALICEunderlyingEvent}: an increase of hadron multiplicity in jets with increasing jet energy
is expected from the evolution of parton cascade with the parton energy for both light and heavy quarks~\cite{Dokshitzer1991}. A decrease of the baseline level with increasing $\pt$ of the associated particles can also be noticed.

Figure~\ref{fig:pp-pPb_Dphi_Compar} shows the $\dphi$ distributions after the subtraction of the baseline, calculated as described
in Section~\ref{sec:fitdescription}. The distributions show
a near-side peak along with a wider and lower peak in the away-side region. The results
obtained for the two collision systems are compatible within the total uncertainties.
According to simulations of pp collisions performed using PYTHIA 6 (Perugia-0, -2010, and -2011 tunes), the
different centre-of-mass energy and the slightly different D-meson rapidity
range of the two measurements should induce variations in the baseline-subtracted azimuthal-correlation distributions smaller
than 7\% in the near- and away-side regions. The same estimate is obtained with POWHEG+PYTHIA simulations including
the EPS09 parametrisation of nuclear PDFs (see Section~\ref{sec:MC}). Such differences are well below the current level of uncertainties. %

A further comparison of the results from pp and p--Pb collisions is done by quantifying the integrals
and the widths of the near-side correlation peaks by fitting the measured distributions as
described in Section~\ref{sec:fitdescription}. The fit results are reported only for the near-side peak parameters
and the baseline because of the poor statistical precision on the fit parameters of the away-side peaks. Figure~\ref{fig:Fittingazimuth}
shows an exemplary fit to the azimuthal-correlation distributions
of D mesons with charged particles with $\ptAssoc>1~\gevc$, for $5<\ptD<8~\gevc$ in pp collisions (left panel)
and for $8<\ptD<16~\gevc$ in p--Pb collisions (right panel). The curves superimposed to the data represent the three terms
of the function defined in Equation~\ref{eq:fitfunction}. 

Within the uncertainties, the fit function describes the measured distributions in all kinematic cases considered, yielding $\chi^{2}/{\rm NDF}$ values close to unity. The evolution of the near-side peak associated yield as a function of the D-meson $\pt$ is reported in Figure~\ref{fig:ppTopPbFitResComp} (top row), for pp and $\pPb$ collisions, for $\ptAssoc>0.3~\gevc$ (left panel) and for the two sub-intervals $0.3<\ptAssoc<1~\gevc$ (middle panel) and $\ptAssoc>1~\gevc$ (right panel). The near-side peak associated yield
exhibits an increasing trend with D-meson $\pt$ and has similar values, within uncertainties, for the softer ($0.3<\ptAssoc<1~\gevc$)
and the harder ($\ptAssoc>1~\gevc$) sub-ranges of $\ptAssoc$ used, in each D-meson $\pt$ interval considered. The values obtained for pp and $\pPb$ collision data are compatible within statistical uncertainties. In the bottom row of the same figure the width
of the near-side Gaussian term ($\sigma_{\rm fit,NS}$) is shown. Although the case with $\ptAssoc>0.3~\gevc$ seems to suggest that
$\sigma_{\rm fit,NS}$ does not strongly depend on D-meson $\pt$ in the range of the measurement, the current level of uncertainty
does not allow for quantification of the dependence of $\sigma_{\rm fit,NS}$ on D-meson and associated
charged particle $\pt$, as well as any potential difference between those values extracted using pp and p--Pb data.
In particular, our approach for baseline calculation (Section~\ref{sec:fitdescription}) guarantees a robust estimate
of the minimum, but the baseline uncertainty and its impact on the associated-yield uncertainty are rather
large (Section~\ref{sec:systematics}). This systematic uncertainty is expected to be significantly reduced in future measurements with larger data samples, where a smaller $\dphi$ range for the baseline calculation could be used.

A $\vtwo$-like modulation of the baseline would introduce a bias in the measurement of the associated yield and peak width and that needs to be taken into account while interpreting the measured quantities in terms of charm-jet properties. In order to get an estimate
of this possible effect, for the p--Pb case the fit was repeated by subtracting from the correlation distribution a $\vtwo$-like modulation
assuming $\vtwo=0.05$ for D mesons and $\vtwo=0.05~(0.1)$ for associated charged particles with $\pt>0.3~(1)~\gev/c$. These values
were chosen on the basis of charged-particle measurements in high-multiplicity p--Pb collisions~\cite{pPbdoubleridgeALICE} and
assuming for D mesons the maximum value predicted in~\cite{beraudoHFv2pPb} for the 20\% most central p--Pb collisions as a test case. With such
assumptions, rather extreme also considering that this measurement is performed without any selection on event multiplicity, $A_{\rm NS}$ varies by
$-10\%$ ($-6\%$) for D mesons with $5<\pt<8~\gev/c$ and for $0.3<\ptAssoc<1~\gev/c$ ($\ptAssoc>1~\gev/c$). The variations on $\sigma_{{\rm fit,NS}}$ and on
the baseline are below 4\% and 1\%, respectively. Significantly smaller modifications result for D mesons with $8<\pt<16~\gev/c$. With the available statistics, the
precision of the measurement is not sufficient to observe or exclude these modifications. %

\begin{figure}[!t]
  \centering
  \begin{minipage}{\linewidth}
    \centering
    \includegraphics[width=\linewidth]{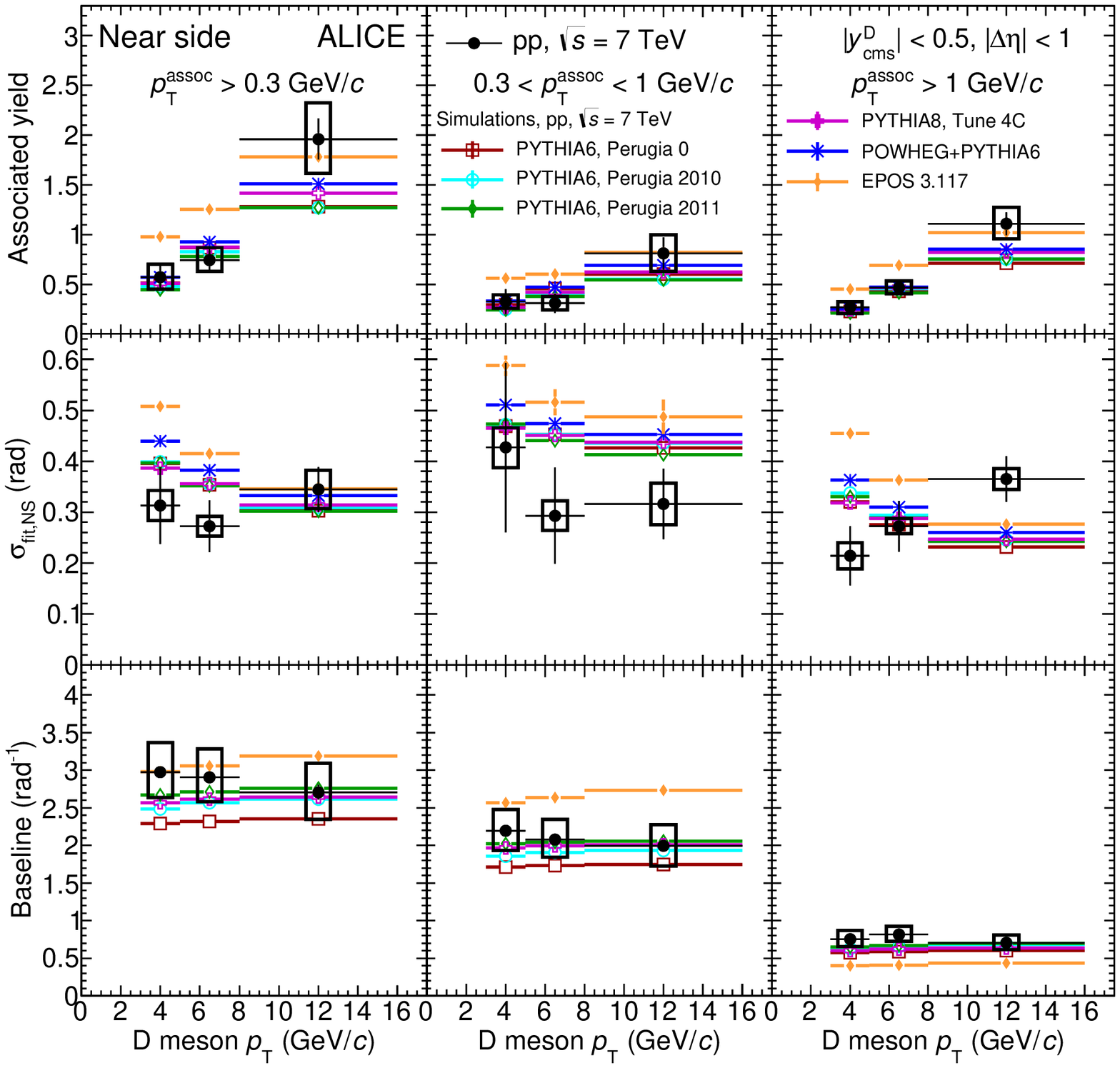}
  \end{minipage}
  \caption{Comparison of near-side peak associated yield (top row), near-side peak width (middle row), and baseline (bottom row) values measured in pp collisions at $\sqrts=7~\tev$ with the expectations from simulations performed with different Monte-Carlo event generators. Statistical and systematic uncertainties are shown as error bars and boxes, respectively.}
  \label{fig:FitParam_CompPPtoMC}
\end{figure}

\begin{figure}[!t]
  \centering
  \begin{minipage}{\linewidth}
    \centering
    \includegraphics[width=\linewidth]{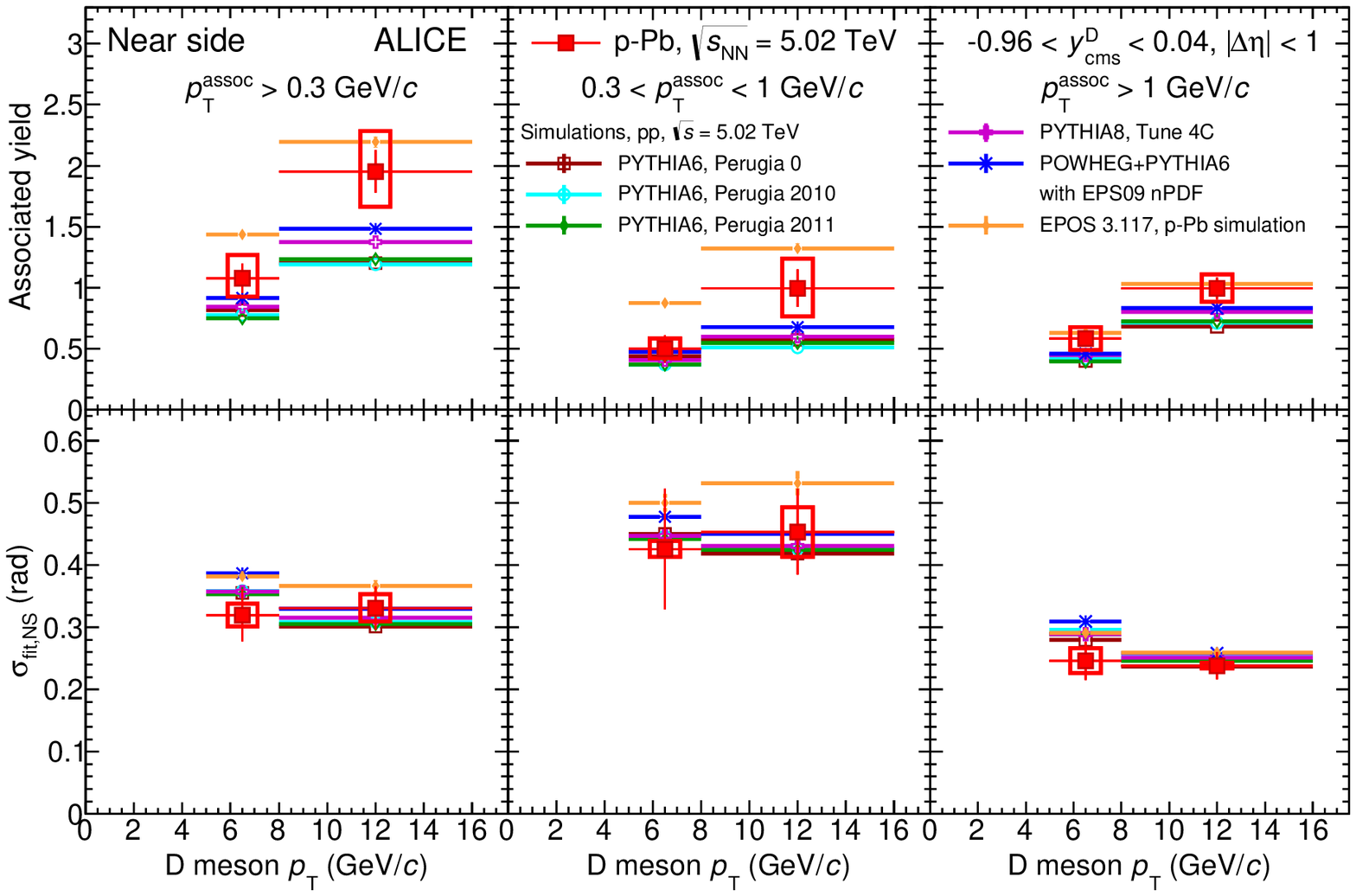}
  \end{minipage}
  \caption{Comparison of near-side peak associated yield (top row) and near-side peak width (bottom row) values measured in p--Pb collisions at $\sqrtsNN=5.02~\tev$ with the expectations from simulations performed with different Monte-Carlo event generators. Statistical and systematic uncertainties are shown as error bars and boxes, respectively.}
  \label{fig:FitParam_CompPPbtoMC}
\end{figure}

Figure~\ref{fig:PYTHIA_Comp} shows the comparison of the averaged azimuthal-correlation distributions measured in
pp collisions with expectations from simulations performed with
PYTHIA, POWHEG+PYTHIA, and EPOS~3 (see Section~\ref{sec:MC}), after the baseline subtraction. The average of the two lowest values of the azimuthal-correlation distribution is used to define the uncertainty related to the baseline definition in Monte-Carlo simulations (see Section~\ref{sec:fitdescription}). This uncertainty is negligible and not displayed in the figures. %
The distributions obtained with the different generators and tunes do not show significant differences in the near side, except from EPOS~3 which tends to
have higher and wider distributions. In the away side, the PYTHIA 6
tunes Perugia~0 and Perugia~2010 tend to have higher correlation values, especially for $\ptAssoc>1~\gevc$, compared to the other
simulation results. Similar considerations hold for EPOS~3 in the case of D mesons with  $8<\pt<16~\gev/c$. The considered Monte-Carlo simulations describe, within the uncertainties, the data in the whole $\dphi$ range. %
The comparison of the associated yield in the near-side peak in data and in simulations is displayed in the top row of Figures~\ref{fig:FitParam_CompPPtoMC} and~\ref{fig:FitParam_CompPPbtoMC}, for pp and p--Pb collisions, respectively. The simulations obtained with EPOS~3 provide a better description of the near-side yields for D mesons with $8<\pt<16~\gev/c$ in both pp and p--Pb collisions. At lower D-meson $\pt$ a better agreement is obtained with PYTHIA and POWHEG+PYTHIA simulations.
The width of the near-side peaks, shown in the second row of the same figures, is better reproduced by the simulations in the case of p--Pb than of pp results. %
The evolution of the baseline value as a function of the D-meson $\pt$ is compared for pp-collision data
to expectations from PYTHIA simulations in the bottom row of Figure~\ref{fig:FitParam_CompPPtoMC} for the
three ranges of $\ptAssoc$ considered in the analysis. The value of the baseline, mainly determined by the event multiplicity, does not
show substantial variations as a function of D-meson $\pt$, as expected
also from PYTHIA and EPOS~3 simulations, which reproduce the observed values within the uncertainties.

\section{Summary}
\label{sec:conclusions}
The first measurements of the azimuthal correlations between D mesons
with charged particles in pp and p--Pb collisions at $\sqrts=7~\tev$ and $\sqrtsNN=5.02~\tev$, respectively, performed
with the ALICE apparatus at the LHC were presented. The $\dphi$
distributions were studied in pp collisions in three different D-meson transverse-momentum intervals, $3<\ptD<5~\gevc$, $5<\ptD<8~\gevc$, and $8<\ptD<16~\gevc$, for
associated charged particles with $\ptAssoc>0.3~\gevc$, and in the two sub-ranges $0.3<\ptAssoc<1~\gevc$ and $\ptAssoc>1~\gevc$. For p--Pb collisions, the results
were reported in two D-meson $\pt$ ranges, $5<\ptD<8~\gevc$, and $8<\ptD<16~\gevc$.
The baseline-subtracted azimuthal-correlation distributions observed in the two collision systems are compatible within uncertainties. The variations expected from the lower nucleon-nucleon centre-of-mass energy of p-Pb collisions and from the slightly different D-meson rapidity ranges used for the p-Pb analysis were studied with simulated pp collisions at the two centre-of-mass energies and are well below the sensitivity of the measurements.

The properties of the near-side correlation peak, sensitive to the characteristics of the jet containing the D meson, were
described in terms of the yield of associated charged particles and peak width, obtained by fitting the $\dphi$ distributions with
a function composed of a constant term, representing the physical minimum of the distribution, and two Gaussian terms
modeling the near- and away-side peaks. The values measured in the two collision systems are compatible
within uncertainties. %

The measured azimuthal distributions, as well as the properties of the correlation peaks, were
compared to expectations from simulations performed with different Monte-Carlo generators. The simulations
reproduce the correlation distributions within uncertainties. %

Considering that the overall uncertainty is dominated by the statistical component, the data collected from pp collisions at $\sqrts=13~\tev$
in the ongoing Run 2 at the LHC will allow for a more precise measurement. In particular, the predicted increase of the cross section
for charm production by more than a factor 2 at $\pt=10~\gevc$ at the higher collision energy~\cite{fonll}, along with the foreseen larger
integrated luminosity, will allow for
a significant reduction of the statistical uncertainty, providing a more quantitative and constraining comparison of the data
with expectations from Monte-Carlo generators. As mentioned in Section~\ref{sec:Results}, with larger data samples a different determination
of the baseline of the azimuthal-correlation distribution will become possible, bringing to a significant reduction of the systematic uncertainty on the measurement of the associated yields. The data that will be collected in next p-Pb collision runs at the LHC may also allow for a study of the evolution of the azimuthal-correlation distribution as a function of the event multiplicity, searching for possible long-range ridge-like structures already observed with angular correlation of light particles.

The results reported in this paper represent a first step towards the measurement of possible modifications concerning the azimuthal
correlation of D mesons with charged particles in Pb--Pb collisions, which has the potential to provide important information
on the charm-quark energy-loss mechanisms in the presence of the medium formed in heavy-ion collisions at LHC energies. Given the same collision energy, the p--Pb results presented in this paper could serve as a reference to study medium effects in Pb--Pb collisions at $\sqrtsNN=5.02~\tev$ collected during the LHC Run 2.

\FloatBarrier

\newenvironment{acknowledgement}{\relax}{\relax}
\begin{acknowledgement}
\section*{Acknowledgements}

The ALICE Collaboration would like to thank all its engineers and technicians for their invaluable contributions to the construction of the experiment and the CERN accelerator teams for the outstanding performance of the LHC complex.
The ALICE Collaboration gratefully acknowledges the resources and support provided by all Grid centres and the Worldwide LHC Computing Grid (WLCG) collaboration.
The ALICE Collaboration acknowledges the following funding agencies for their support in building and
running the ALICE detector:
State Committee of Science,  World Federation of Scientists (WFS)
and Swiss Fonds Kidagan, Armenia;
Conselho Nacional de Desenvolvimento Cient\'{\i}fico e Tecnol\'{o}gico (CNPq), Financiadora de Estudos e Projetos (FINEP),
Funda\c{c}\~{a}o de Amparo \`{a} Pesquisa do Estado de S\~{a}o Paulo (FAPESP);
Ministry of Science \& Technology of China (MSTC), National Natural Science Foundation of China (NSFC) and Ministry of Education of China (MOEC)";
Ministry of Science, Education and Sports of Croatia and  Unity through Knowledge Fund, Croatia;
Ministry of Education and Youth of the Czech Republic;
Danish Natural Science Research Council, the Carlsberg Foundation and the Danish National Research Foundation;
The European Research Council under the European Community's Seventh Framework Programme;
Helsinki Institute of Physics and the Academy of Finland;
French CNRS-IN2P3, the `Region Pays de Loire', `Region Alsace', `Region Auvergne' and CEA, France;
German Bundesministerium fur Bildung, Wissenschaft, Forschung und Technologie (BMBF) and the Helmholtz Association;
General Secretariat for Research and Technology, Ministry of Development, Greece;
National Research, Development and Innovation Office (NKFIH), Hungary;
Council of Scientific and Industrial Research (CSIR), New Delhi;
Department of Atomic Energy and Department of Science and Technology of the Government of India;
Istituto Nazionale di Fisica Nucleare (INFN) and Centro Fermi - Museo Storico della Fisica e Centro Studi e Ricerche ``Enrico Fermi'', Italy;
Japan Society for the Promotion of Science (JSPS) KAKENHI and MEXT, Japan;
National Research Foundation of Korea (NRF);
Consejo Nacional de Cienca y Tecnologia (CONACYT), Direccion General de Asuntos del Personal Academico(DGAPA), M\'{e}xico, Amerique Latine Formation academique - 
European Commission~(ALFA-EC) and the EPLANET Program~(European Particle Physics Latin American Network);
Stichting voor Fundamenteel Onderzoek der Materie (FOM) and the Nederlandse Organisatie voor Wetenschappelijk Onderzoek (NWO), Netherlands;
Research Council of Norway (NFR);
Pontificia Universidad Cat\'{o}lica del Per\'{u};
National Science Centre, Poland;
Ministry of National Education/Institute for Atomic Physics and National Council of Scientific Research in Higher Education~(CNCSI-UEFISCDI), Romania;
Joint Institute for Nuclear Research, Dubna;
Ministry of Education and Science of Russian Federation, Russian Academy of Sciences, Russian Federal Agency of Atomic Energy, Russian Federal Agency for Science and Innovations and The Russian Foundation for Basic Research;
Ministry of Education of Slovakia;
Department of Science and Technology, South Africa;
Centro de Investigaciones Energeticas, Medioambientales y Tecnologicas (CIEMAT), E-Infrastructure shared between Europe and Latin America (EELA), 
Ministerio de Econom\'{i}a y Competitividad (MINECO) of Spain, Xunta de Galicia (Conseller\'{\i}a de Educaci\'{o}n),
Centro de Aplicaciones Tecnológicas y Desarrollo Nuclear (CEA\-DEN), Cubaenerg\'{\i}a, Cuba, and IAEA (International Atomic Energy Agency);
Swedish Research Council (VR) and Knut $\&$ Alice Wallenberg Foundation (KAW);
National Science and Technology Development Agency (NSDTA), Suranaree University of Technology (SUT) and Office of the Higher Education Commission under NRU project of Thailand;
Ukraine Ministry of Education and Science;
United Kingdom Science and Technology Facilities Council (STFC);
The United States Department of Energy, the United States National Science Foundation, the State of Texas, and the State of Ohio.

\end{acknowledgement}

\bibliographystyle{utphys}   
\bibliography{biblio}

\newpage
\appendix
\section{The ALICE Collaboration}
\label{app:collab}



\begingroup
\small
\begin{flushleft}
J.~Adam$^\textrm{\scriptsize 39}$,
D.~Adamov\'{a}$^\textrm{\scriptsize 85}$,
M.M.~Aggarwal$^\textrm{\scriptsize 89}$,
G.~Aglieri Rinella$^\textrm{\scriptsize 35}$,
M.~Agnello$^\textrm{\scriptsize 112}$\textsuperscript{,}$^\textrm{\scriptsize 31}$,
N.~Agrawal$^\textrm{\scriptsize 48}$,
Z.~Ahammed$^\textrm{\scriptsize 136}$,
S.~Ahmad$^\textrm{\scriptsize 18}$,
S.U.~Ahn$^\textrm{\scriptsize 69}$,
S.~Aiola$^\textrm{\scriptsize 140}$
% ...
\renewcommand\labelenumi{\textsuperscript{\theenumi}~}

\section*{Affiliation notes}
\renewcommand\theenumi{\roman{enumi}}
\begin{Authlist}
\item \Adef{0}Deceased
\item \Adef{idp1824064}{Also at: Georgia State University, Atlanta, Georgia, United States}
\item \Adef{idp3239488}{Also at: Also at Department of Applied Physics, Aligarh Muslim University, Aligarh, India}
\item \Adef{idp3964800}{Also at: M.V. Lomonosov Moscow State University, D.V. Skobeltsyn Institute of Nuclear, Physics, Moscow, Russia}
\end{Authlist}

\section*{Collaboration Institutes}
\renewcommand\theenumi{\arabic{enumi}~}

$^{1}$A.I. Alikhanyan National Science Laboratory (Yerevan Physics Institute) Foundation, Yerevan, Armenia
\\
$^{2}$Benem\'{e}rita Universidad Aut\'{o}noma de Puebla, Puebla, Mexico
\\
$^{3}$Bogolyubov Institute for Theoretical Physics, Kiev, Ukraine
\\
$^{4}$Bose Institute, Department of Physics 
and Centre for Astroparticle Physics and Space Science (CAPSS), Kolkata, India
\\
$^{5}$Budker Institute for Nuclear Physics, Novosibirsk, Russia
\\
$^{6}$California Polytechnic State University, San Luis Obispo, California, United States
\\
$^{7}$Central China Normal University, Wuhan, China
\\
$^{8}$Centre de Calcul de l'IN2P3, Villeurbanne, Lyon, France
\\
$^{9}$Centro de Aplicaciones Tecnol\'{o}gicas y Desarrollo Nuclear (CEADEN), Havana, Cuba
\\
$^{10}$Centro de Investigaciones Energ\'{e}ticas Medioambientales y Tecnol\'{o}gicas (CIEMAT), Madrid, Spain
\\
$^{11}$Centro de Investigaci\'{o}n y de Estudios Avanzados (CINVESTAV), Mexico City and M\'{e}rida, Mexico
\\
$^{12}$Centro Fermi - Museo Storico della Fisica e Centro Studi e Ricerche ``Enrico Fermi', Rome, Italy
\\
$^{13}$Chicago State University, Chicago, Illinois, United States
\\
$^{14}$China Institute of Atomic Energy, Beijing, China
\\
$^{15}$Commissariat \`{a} l'Energie Atomique, IRFU, Saclay, France
\\
$^{16}$COMSATS Institute of Information Technology (CIIT), Islamabad, Pakistan
\\
$^{17}$Departamento de F\'{\i}sica de Part\'{\i}culas and IGFAE, Universidad de Santiago de Compostela, Santiago de Compostela, Spain
\\
$^{18}$Department of Physics, Aligarh Muslim University, Aligarh, India
\\
$^{19}$Department of Physics, Ohio State University, Columbus, Ohio, United States
\\
$^{20}$Department of Physics, Sejong University, Seoul, South Korea
\\
$^{21}$Department of Physics, University of Oslo, Oslo, Norway
\\
$^{22}$Department of Physics and Technology, University of Bergen, Bergen, Norway
\\
$^{23}$Dipartimento di Fisica dell'Universit\`{a} 'La Sapienza'
and Sezione INFN, Rome, Italy
\\
$^{24}$Dipartimento di Fisica dell'Universit\`{a}
and Sezione INFN, Cagliari, Italy
\\
$^{25}$Dipartimento di Fisica dell'Universit\`{a}
and Sezione INFN, Trieste, Italy
\\
$^{26}$Dipartimento di Fisica dell'Universit\`{a}
and Sezione INFN, Turin, Italy
\\
$^{27}$Dipartimento di Fisica e Astronomia dell'Universit\`{a}
and Sezione INFN, Bologna, Italy
\\
$^{28}$Dipartimento di Fisica e Astronomia dell'Universit\`{a}
and Sezione INFN, Catania, Italy
\\
$^{29}$Dipartimento di Fisica e Astronomia dell'Universit\`{a}
and Sezione INFN, Padova, Italy
\\
$^{30}$Dipartimento di Fisica `E.R.~Caianiello' dell'Universit\`{a}
and Gruppo Collegato INFN, Salerno, Italy
\\
$^{31}$Dipartimento DISAT del Politecnico and Sezione INFN, Turin, Italy
\\
$^{32}$Dipartimento di Scienze e Innovazione Tecnologica dell'Universit\`{a} del Piemonte Orientale and INFN Sezione di Torino, Alessandria, Italy
\\
$^{33}$Dipartimento Interateneo di Fisica `M.~Merlin'
and Sezione INFN, Bari, Italy
\\
$^{34}$Division of Experimental High Energy Physics, University of Lund, Lund, Sweden
\\
$^{35}$European Organization for Nuclear Research (CERN), Geneva, Switzerland
\\
$^{36}$Excellence Cluster Universe, Technische Universit\"{a}t M\"{u}nchen, Munich, Germany
\\
$^{37}$Faculty of Engineering, Bergen University College, Bergen, Norway
\\
$^{38}$Faculty of Mathematics, Physics and Informatics, Comenius University, Bratislava, Slovakia
\\
$^{39}$Faculty of Nuclear Sciences and Physical Engineering, Czech Technical University in Prague, Prague, Czech Republic
\\
$^{40}$Faculty of Science, P.J.~\v{S}af\'{a}rik University, Ko\v{s}ice, Slovakia
\\
$^{41}$Faculty of Technology, Buskerud and Vestfold University College, Tonsberg, Norway
\\
$^{42}$Frankfurt Institute for Advanced Studies, Johann Wolfgang Goethe-Universit\"{a}t Frankfurt, Frankfurt, Germany
\\
$^{43}$Gangneung-Wonju National University, Gangneung, South Korea
\\
$^{44}$Gauhati University, Department of Physics, Guwahati, India
\\
$^{45}$Helmholtz-Institut f\"{u}r Strahlen- und Kernphysik, Rheinische Friedrich-Wilhelms-Universit\"{a}t Bonn, Bonn, Germany
\\
$^{46}$Helsinki Institute of Physics (HIP), Helsinki, Finland
\\
$^{47}$Hiroshima University, Hiroshima, Japan
\\
$^{48}$Indian Institute of Technology Bombay (IIT), Mumbai, India
\\
$^{49}$Indian Institute of Technology Indore, Indore, India
\\
$^{50}$Indonesian Institute of Sciences, Jakarta, Indonesia
\\
$^{51}$Inha University, Incheon, South Korea
\\
$^{52}$Institut de Physique Nucl\'eaire d'Orsay (IPNO), Universit\'e Paris-Sud, CNRS-IN2P3, Orsay, France
\\
$^{53}$Institute for Nuclear Research, Academy of Sciences, Moscow, Russia
\\
$^{54}$Institute for Subatomic Physics of Utrecht University, Utrecht, Netherlands
\\
$^{55}$Institute for Theoretical and Experimental Physics, Moscow, Russia
\\
$^{56}$Institute of Experimental Physics, Slovak Academy of Sciences, Ko\v{s}ice, Slovakia
\\
$^{57}$Institute of Physics, Academy of Sciences of the Czech Republic, Prague, Czech Republic
\\
$^{58}$Institute of Physics, Bhubaneswar, India
\\
$^{59}$Institute of Space Science (ISS), Bucharest, Romania
\\
$^{60}$Institut f\"{u}r Informatik, Johann Wolfgang Goethe-Universit\"{a}t Frankfurt, Frankfurt, Germany
\\
$^{61}$Institut f\"{u}r Kernphysik, Johann Wolfgang Goethe-Universit\"{a}t Frankfurt, Frankfurt, Germany
\\
$^{62}$Institut f\"{u}r Kernphysik, Westf\"{a}lische Wilhelms-Universit\"{a}t M\"{u}nster, M\"{u}nster, Germany
\\
$^{63}$Instituto de Ciencias Nucleares, Universidad Nacional Aut\'{o}noma de M\'{e}xico, Mexico City, Mexico
\\
$^{64}$Instituto de F\'{\i}sica, Universidad Nacional Aut\'{o}noma de M\'{e}xico, Mexico City, Mexico
\\
$^{65}$Institut Pluridisciplinaire Hubert Curien (IPHC), Universit\'{e} de Strasbourg, CNRS-IN2P3, Strasbourg, France
\\
$^{66}$iThemba LABS, National Research Foundation, Somerset West, South Africa
\\
$^{67}$Joint Institute for Nuclear Research (JINR), Dubna, Russia
\\
$^{68}$Konkuk University, Seoul, South Korea
\\
$^{69}$Korea Institute of Science and Technology Information, Daejeon, South Korea
\\
$^{70}$KTO Karatay University, Konya, Turkey
\\
$^{71}$Laboratoire de Physique Corpusculaire (LPC), Clermont Universit\'{e}, Universit\'{e} Blaise Pascal, CNRS--IN2P3, Clermont-Ferrand, France
\\
$^{72}$Laboratoire de Physique Subatomique et de Cosmologie, Universit\'{e} Grenoble-Alpes, CNRS-IN2P3, Grenoble, France
\\
$^{73}$Laboratori Nazionali di Frascati, INFN, Frascati, Italy
\\
$^{74}$Laboratori Nazionali di Legnaro, INFN, Legnaro, Italy
\\
$^{75}$Lawrence Berkeley National Laboratory, Berkeley, California, United States
\\
$^{76}$Moscow Engineering Physics Institute, Moscow, Russia
\\
$^{77}$Nagasaki Institute of Applied Science, Nagasaki, Japan
\\
$^{78}$National Centre for Nuclear Studies, Warsaw, Poland
\\
$^{79}$National Institute for Physics and Nuclear Engineering, Bucharest, Romania
\\
$^{80}$National Institute of Science Education and Research, Bhubaneswar, India
\\
$^{81}$National Research Centre Kurchatov Institute, Moscow, Russia
\\
$^{82}$Niels Bohr Institute, University of Copenhagen, Copenhagen, Denmark
\\
$^{83}$Nikhef, Nationaal instituut voor subatomaire fysica, Amsterdam, Netherlands
\\
$^{84}$Nuclear Physics Group, STFC Daresbury Laboratory, Daresbury, United Kingdom
\\
$^{85}$Nuclear Physics Institute, Academy of Sciences of the Czech Republic, \v{R}e\v{z} u Prahy, Czech Republic
\\
$^{86}$Oak Ridge National Laboratory, Oak Ridge, Tennessee, United States
\\
$^{87}$Petersburg Nuclear Physics Institute, Gatchina, Russia
\\
$^{88}$Physics Department, Creighton University, Omaha, Nebraska, United States
\\
$^{89}$Physics Department, Panjab University, Chandigarh, India
\\
$^{90}$Physics Department, University of Athens, Athens, Greece
\\
$^{91}$Physics Department, University of Cape Town, Cape Town, South Africa
\\
$^{92}$Physics Department, University of Jammu, Jammu, India
\\
$^{93}$Physics Department, University of Rajasthan, Jaipur, India
\\
$^{94}$Physikalisches Institut, Eberhard Karls Universit\"{a}t T\"{u}bingen, T\"{u}bingen, Germany
\\
$^{95}$Physikalisches Institut, Ruprecht-Karls-Universit\"{a}t Heidelberg, Heidelberg, Germany
\\
$^{96}$Physik Department, Technische Universit\"{a}t M\"{u}nchen, Munich, Germany
\\
$^{97}$Purdue University, West Lafayette, Indiana, United States
\\
$^{98}$Pusan National University, Pusan, South Korea
\\
$^{99}$Research Division and ExtreMe Matter Institute EMMI, GSI Helmholtzzentrum f\"ur Schwerionenforschung, Darmstadt, Germany
\\
$^{100}$Rudjer Bo\v{s}kovi\'{c} Institute, Zagreb, Croatia
\\
$^{101}$Russian Federal Nuclear Center (VNIIEF), Sarov, Russia
\\
$^{102}$Saha Institute of Nuclear Physics, Kolkata, India
\\
$^{103}$School of Physics and Astronomy, University of Birmingham, Birmingham, United Kingdom
\\
$^{104}$Secci\'{o}n F\'{\i}sica, Departamento de Ciencias, Pontificia Universidad Cat\'{o}lica del Per\'{u}, Lima, Peru
\\
$^{105}$Sezione INFN, Bari, Italy
\\
$^{106}$Sezione INFN, Bologna, Italy
\\
$^{107}$Sezione INFN, Cagliari, Italy
\\
$^{108}$Sezione INFN, Catania, Italy
\\
$^{109}$Sezione INFN, Padova, Italy
\\
$^{110}$Sezione INFN, Rome, Italy
\\
$^{111}$Sezione INFN, Trieste, Italy
\\
$^{112}$Sezione INFN, Turin, Italy
\\
$^{113}$SSC IHEP of NRC Kurchatov institute, Protvino, Russia
\\
$^{114}$Stefan Meyer Institut f\"{u}r Subatomare Physik (SMI), Vienna, Austria
\\
$^{115}$SUBATECH, Ecole des Mines de Nantes, Universit\'{e} de Nantes, CNRS-IN2P3, Nantes, France
\\
$^{116}$Suranaree University of Technology, Nakhon Ratchasima, Thailand
\\
$^{117}$Technical University of Ko\v{s}ice, Ko\v{s}ice, Slovakia
\\
$^{118}$Technical University of Split FESB, Split, Croatia
\\
$^{119}$The Henryk Niewodniczanski Institute of Nuclear Physics, Polish Academy of Sciences, Cracow, Poland
\\
$^{120}$The University of Texas at Austin, Physics Department, Austin, Texas, United States
\\
$^{121}$Universidad Aut\'{o}noma de Sinaloa, Culiac\'{a}n, Mexico
\\
$^{122}$Universidade de S\~{a}o Paulo (USP), S\~{a}o Paulo, Brazil
\\
$^{123}$Universidade Estadual de Campinas (UNICAMP), Campinas, Brazil
\\
$^{124}$Universidade Federal do ABC, Santo Andre, Brazil
\\
$^{125}$University of Houston, Houston, Texas, United States
\\
$^{126}$University of Jyv\"{a}skyl\"{a}, Jyv\"{a}skyl\"{a}, Finland
\\
$^{127}$University of Liverpool, Liverpool, United Kingdom
\\
$^{128}$University of Tennessee, Knoxville, Tennessee, United States
\\
$^{129}$University of the Witwatersrand, Johannesburg, South Africa
\\
$^{130}$University of Tokyo, Tokyo, Japan
\\
$^{131}$University of Tsukuba, Tsukuba, Japan
\\
$^{132}$University of Zagreb, Zagreb, Croatia
\\
$^{133}$Universit\'{e} de Lyon, Universit\'{e} Lyon 1, CNRS/IN2P3, IPN-Lyon, Villeurbanne, Lyon, France
\\
$^{134}$Universit\`{a} di Brescia, Brescia, Italy
\\
$^{135}$V.~Fock Institute for Physics, St. Petersburg State University, St. Petersburg, Russia
\\
$^{136}$Variable Energy Cyclotron Centre, Kolkata, India
\\
$^{137}$Warsaw University of Technology, Warsaw, Poland
\\
$^{138}$Wayne State University, Detroit, Michigan, United States
\\
$^{139}$Wigner Research Centre for Physics, Hungarian Academy of Sciences, Budapest, Hungary
\\
$^{140}$Yale University, New Haven, Connecticut, United States
\\
$^{141}$Yonsei University, Seoul, South Korea
\\
$^{142}$Zentrum f\"{u}r Technologietransfer und Telekommunikation (ZTT), Fachhochschule Worms, Worms, Germany
\endgroup

\end{document}